\documentclass[12pt]{article}
\usepackage{a4wide,graphics,graphicx,amsmath,amssymb,cite,nicefrac}
\usepackage[verbose]{wrapfig}
\usepackage{authblk,hyperref}
\usepackage{url,amsfonts}
\usepackage[makeroom]{cancel}
\catcode`@=11 \@addtoreset{equation}{section} \catcode`@=12

\usepackage[english]{babel}
\usepackage{mathrsfs}
\usepackage{amsmath}
\usepackage{amsfonts}
\usepackage{amssymb}
\usepackage{makeidx}
\usepackage{graphicx}
\usepackage{physics}
\usepackage{float} %Include figure filesusepackage{graphicx} %Include figure files
\usepackage{subfloat}
\usepackage{multicol}
\usepackage{subcaption}
\usepackage{xcolor}
\def\ie{\begin{equation}\begin{aligned}}
\def\fe{\end{aligned}\end{equation}}

\usepackage{tikz}

\def\ie{\begin{equation}\begin{aligned}}
\def\fe{\end{aligned}\end{equation}}

\newcommand{\cA}{\mathcal A}

\newcommand{\cF}{\mathcal F}

\DeclareFontEncoding{LS1}{}{}
\DeclareFontSubstitution{LS1}{stix}{m}{n}
\DeclareSymbolFont{stixsymbols}{LS1}{stixscr}{m}{n}
\SetSymbolFont{stixsymbols}{bold}{LS1}{stixscr}{b}{n}
\DeclareMathSymbol{\kay}{\mathalpha}{stixsymbols}{"6B}
\DeclareMathSymbol{\hay}{\mathalpha}{stixsymbols}{"68}

\DeclareMathAlphabet{\mathdsl}{U}{bbm}{m}{sl}

\begin{document}

\begin{titlepage}%1
\begin{center}
%\hfill DFPD/2017/TH/\\
\vskip 1.0cm
{\bf \Large The Yang--Baxter Sigma Model  from Twistor Space}
\vskip 1cm
{Meer Ashwinkumar$^{a}$ and Jitendra Pal$^{b}$
} \vskip 0.05in 
\end{center}
\vskip 0.5cm 
\begin{center}
{\small{ \textit{$^{a}$Institute for Theoretical Physics, Albert Einstein Center for Fundamental Physics,\\  University of Bern,
Sidlerstrasse 5, CH-3012 Bern, Switzerland}}}\\
\textit{$^b$Wilczek Quantum Center, Shanghai Institute for Advanced Studies, University of Science and Technology of China,
Shanghai 201315, China}
\end{center}
\begin{center}
    E-mail :  \href{mailto:meer.ashwinkumar@unibe.ch}{meer.ashwinkumar@unibe.ch}, \href{mailto:jeetupal007@gmail.com}{jeetupal007@gmail.com}
\end{center}

\vskip 1cm

\begin{%
center} {\bf Abstract}\\[3ex]\end{center}%
We derive a novel two-field four-dimensional integrable field theory (IFT) from 6d holomorphic Chern--Simons theory on twistor space. The four-dimensional IFT depends on a skew-symmetric linear operator acting on a Lie algebra, and when this operator is specialised to a solution of the modified classical Yang--Baxter equation, the IFT develops a semi-local symmetry associated with this solution. The resulting 4d analogue of the Yang--Baxter sigma model is related by symmetry reduction to the well-known 2d Yang--Baxter sigma model. An important implication that we find is the embedding of the equations of motion of the 2d Yang--Baxter sigma model in the anti-self-dual Yang--Mills equations.  The 6d Chern--Simons theory on twistor space can alternatively be symmetry reduced to a 4d Chern--Simons theory configuration with disorder surface defects. The latter realises the Yang--Baxter sigma model, implying a ``diamond" for the Yang--Baxter sigma model obtained from twistor space.  We also show that the homogeneous 2d Yang--Baxter sigma model can be derived from a limit of our setup.

%\today

%\end{center}%

%\noindent

%
\vfill

%\July 2008

\end{titlepage}

\newpage % \setcounter{page}{1} \numberwithin{equation}{section}
\tableofcontents
\section{Introduction}

Two-dimensional integrable field theories serve as valuable laboratories for exploring the structure of quantum field theory. Many such models, including the principal chiral model, are asymptotically free, and their constrained dynamics offer insight into the behaviour of more realistic asymptotically free theories, such as four-dimensional Yang–Mills theory.

Four-dimensional Chern--Simons (CS) theory has emerged as a unifying framework for the study of two-dimensional integrable systems \cite{Costello:2013zra, Costello:2017dso, Costello:2018gyb, Costello:2021zcl}, encompassing integrable lattice models governed by Yangian symmetry, as well as integrable field theories \cite{Costello:2019tri, Ashwinkumar:2023zbu, Levine:2023wvt, Schmidtt:2023slc, Schmidtt:2025zsb}, with further connections to integrable structures arising in supersymmetric field theories \cite{Ashwinkumar:2018tmm, Costello:2018txb, Ashwinkumar:2019mtj} and holographically dual theories \cite{Ishtiaque:2018str, Ashwinkumar:2020gxt}.  In particular, Costello and Yamazaki \cite{Costello:2019tri} showed that a broad class of two-dimensional integrable field theories can be obtained from four-dimensional Chern–Simons theory through suitable choices of a meromorphic one-form, $\omega$, and boundary conditions. The action takes the form
\ie 
S=\frac{1}{2\pi i}\int_{\Sigma \times C} \omega \wedge \textrm{CS}(A), 
\fe 
where $\textrm{CS}(A)$ is the Chern--Simons 3-form, $\Sigma$ is the worldsheet of the integrable field theory and $C$ is a Riemann surface.  
This perspective on integrable field theories has since led to numerous generalisations and further developments; see \cite{ Sakamoto:2025hwi, Fukushima:2025tlj, Fukushima:2026gan, Benini:2026pwa, Stedman:2026awg, Bittleston:2026tdr, Cole:2025zmq, Hamidi:2025sgg} for recent progress. 

Another well-known organising principle for 2d integrable field theories is the anti-self-dual Yang--Mills (ASDYM) equations \cite{mason1996integrability}. Various integrable systems such as the Korteweg-de Vries (KdV) equation, nonlinear Schroedinger equation, and Toda field equation can be derived from the 
ASDYM equations via different symmetry reductions. 

The 
relationship between the two frameworks of 4d CS and the ASDYM equations can be understood  through the lens of 6d holomorphic Chern--Simons theory on twistor space.
This was first elucidated in the work of Bittleston and Skinner \cite{Bittleston:2020hfv} and Penna \cite{Penna:2020uky}, where the embedding of the principal chiral model in  6d holomorphic CS was elucidated. Further work in this direction clarified the embedding of integrable deformations of the principal chiral model, as well as coset models and non-abelian T-duals \cite{He:2021xoo, Cole:2023umd, Cole:2024ess, Chatzis:2025lly}.

In this work, we shall present an embedding of the Yang--Baxter deformation of the principal chiral model (also known as the Yang--Baxter sigma model) in 6d holomorphic CS on twistor space. 
The Yang--Baxter (YB) sigma model is a well-known example of a deformation of the principal chiral model (PCM) that preserves integrability, that originates in the work of Klimcik \cite{Klimcik:2002zj,Klimcik:2008eq}; see also \cite{Matsumoto:2015jja}. It depends explicitly on a solution of the classical Yang--Baxter equation or its modified version. The deformation can be interpreted as deformation of the geometry of the Lie group target space of the principal chiral model, along with the inclusion of a nontrivial $B$-field. 

The YB sigma model has proven to be a canonical example of an integrable deformation of the PCM, exhibiting rich dynamical structure. Notably, an analogue of the YB deformation also exists for coset models, a prime example of which is the YB deformation of the $AdS_5 \times S^5$ superstring \cite{Delduc:2013qra, Kawaguchi:2014qwa , Delduc:2014kha, vanTongeren:2015soa } (derived from 4d Chern-Simons theory in \cite{Fukushima:2020dcp}). The existence of the latter shows that integrability in the AdS/CFT correspondence survives a deformation of the $AdS_5 \times S^5$ geometry. Moreover, integrability constrains the quantum dynamics of the YB model strongly enough such that we can derive its nonperturbative physics systematically \cite{Schepers:2020ehn, Ashwinkumar:2025wad}.

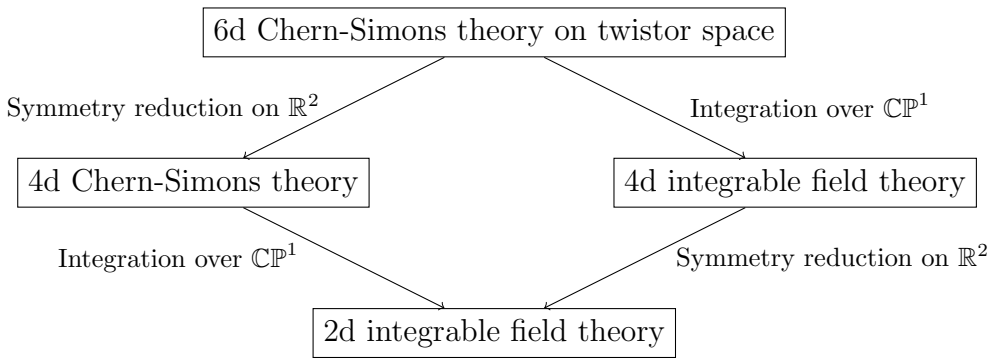
\begin{figure}[h]
\begin{center}
\begin{tikzpicture}[scale=1]
\path (0,0) node[rectangle,draw](2D)  {2d integrable field theory}
(0,4) node[rectangle,draw] (6D) {6d Chern-Simons theory on twistor space}
(-4,2) node[rectangle,draw] (4DCS) {4d Chern-Simons theory}
(4,2) node[rectangle,draw] (4DI) {4d integrable field theory};
\draw[->] (6D) -- (4DCS)
 node[pos=0.5, xshift=-2.4cm]{{\footnotesize Symmetry reduction on $\mathbb{R}^2$}};
\draw[->] (6D) -- (4DI)
node[midway, xshift=2.2cm]{{\footnotesize Integration over $\mathbb{CP}^1$}};
\draw[->] (4DCS) -- (2D)
node[pos=0.5,xshift=-2.2cm]{{\footnotesize Integration over $\mathbb{CP}^1$ }};
\draw[->] (4DI) -- (2D)
node[pos=0.5,xshift=2.5cm]{{\footnotesize Symmetry reduction on $\mathbb{R}^2$}};
\end{tikzpicture}
\end{center}
\caption{Relationship between 4d Chern-Simons theory and 6d Chern-Simons theory on twistor space, as well as 4d and 2d integrable field theories, forming a ``diamond".}
\label{twistor}
\end{figure}

We recall that the derivation of the Yang--Baxter sigma model from 4d Chern--Simons theory is well-known from the work of Delduc et al. \cite{Delduc:2019whp} (an alternative derivation is found in \cite{Fukushima:2020kta}). In this derivation, the datum of the solution of the classical Yang--Baxter equation enters via boundary conditions in the Chern--Simons theory. In this work, we shall identify this $4\textrm{d} \rightarrow 2\textrm{d}$ reduction with an edge in a ``diamond" of reductions that start with 6d holomorphic CS on twistor space, depicted in Figure \ref{twistor}. We shall derive a novel 4d IFT that is an analogue of the Yang--Baxter sigma model from this 6d holomorphic CS theory. This 4d IFT can moreover be symmetry reduced to the 2d Yang--Baxter sigma model, completing the diamond. In the process, we shall show that the equations of motion of the Yang--Baxter sigma model can be embedded in the 4d ASDYM equations.

The remainder of this article is organised as follows. In Section 2,
we shall show that 6d holomorphic CS with boundary conditions that depend on a linear, antisymmetric operator reduces to a two-field 4d integrable field theory, also referred to as an $\textrm{IFT}_4$. We further describe how the equations of motion of this theory are equivalent to the anti-self-dual Yang--Mills equations.   
In Section 3, we describe the symmetry reduction of our 6d CS action and boundary conditions to a 4d CS action with disorder defects and compatible boundary conditions. 
In Section 4,
we describe the features of the 4d IFT  when the linear operator is specialised to a solution of the modified classical Yang-Baxter equation. We refer to the 4d IFT with such a specialisation as $\textrm{IFT}_4^{\textrm{YB}}$, or the 4d Yang--Baxter sigma model.
In Section 5, we show that $\textrm{IFT}_4$ can be related via a symmetry reduction to a two-field 2d theory, that can be further identified with the (inhomogeneous) Yang--Baxter sigma model in the presence of a gauge symmetry. 
In Section 6, we explain how the aforementioned two-field 2d theory arises from a 4d CS setup, completing a diamond of reductions. In Sections 7 and 8, we describe how to derive the Yang--Baxter sigma model associated with the homogeneous classical Yang--Baxter equation from a limit of our setup. Finally, in Section 9, we summarize our results and discuss possible future directions.
\newline
\newline
\noindent
\textbf{Acknowledgements }: We would like to thank Ben Hoare for many helpful conversations and comments on the manuscript.  We would also like to thank Sujay K. Ashok, Roland Bittleston, and Matthias Blau for helpful discussions. We also would like to thank the anonymous referee for helpful feedback and suggestions.
Meer Ashwinkumar is supported in part by NCCR SwissMAP - ``The Mathematics of Physics" of the Swiss National Science Foundation. 

\section{4d IFT from 6d CS on Twistor Space}\label{Sec2}

We begin with the holomorphic 6d  Chern-Simons theory on twistor space, whose action is given by 

\begin{equation}
S_{\mathrm{hCS}_6}
  = \frac{1}{2\pi i} \int_{\mathbb{PT}} 
    \Omega \wedge \mathrm{Tr}\left(
      \cA \wedge \bar{\partial} \cA 
      + \frac{2}{3} \cA \wedge \cA \wedge \cA
    \right),
    \label{6daction}
\end{equation}
with the
$(3,0)$-form given by
\begin{equation}\label{twist function}
\Omega = \frac{1}{2}\,\Phi\, e^{0} \wedge e^{A} \wedge e_{A},
\qquad
\Phi = \frac{K}{
  \langle \pi \alpha \rangle
  \langle \pi \tilde{\alpha} \rangle
  \langle \pi \beta \rangle^{2}
 } .
\end{equation}
Our conventions for twistor space are collected in Appendix \ref{ap A}.
The gauge field in the 
$(0,1)$-form basis is given by 
\begin{equation}
    \cA = \cA_{0}\,\bar{e}^{0} + \cA_{A}\,\bar{e}^{A},
\end{equation}
and the variation of the holomorphic Chern–Simons action is
\begin{equation}
\delta S = \frac{1}{2\pi i} \int_{\mathbb{PT}} 
\Big[ 2\,\Omega \wedge \mathrm{Tr}(\delta \cA \wedge \cF^{0,2}) 
+ \bar{\partial}\Omega \wedge \mathrm{Tr}(\cA \wedge \delta \cA) \Big].
\end{equation}
For the action to be stationary for arbitrary $\delta \cA$, one requires that
the bulk equation of motion 
\ie 	\cF^{0,2} = \bar{\partial}\cA + \cA \wedge \cA = 0,\fe and the boundary equation of motion
\begin{align}\label{defect term}
    	0 = \int_{\mathbb{PT}} \bar{\partial}\Omega \wedge \mathrm{Tr}(\cA \wedge \delta \cA),
\end{align}
hold. 
Since all singularities of $\Omega$ are encoded in the scalar prefactor $\Phi$, the $\bar\partial$-operator localises the integral to the poles on the $\mathbb{CP}^1$ fibre. From \eqref{twist function},  $\Omega$ has poles on $\mathbb{CP}^1$ at $\pi = \alpha$ (simple pole), $\pi = \tilde{\alpha}$ (simple pole), $\pi = \beta$ (double pole). We consider the Dirichlet boundary conditions $\cA_A\big|_{\pi=\beta}=0$ for the second order pole. The resulting residues yield pole-localised terms proportional to
$\int_{E^4}\mathrm{vol}_4\,\epsilon^{AB}\mathrm{Tr}(\cA_A \delta \cA_B)$ evaluated at the respective poles. Using the antisymmetry of the spinor bracket,
$\langle \tilde\alpha \alpha \rangle = - \langle \alpha \tilde\alpha \rangle$,
the boundary equation of motion can be stated as 
\begin{equation}
\frac{1}{\langle \alpha \tilde{\alpha}\rangle \langle \alpha \beta\rangle^{2}}
\int_{\mathbb{E}^{4}} \mathrm{vol}_{4}\,\epsilon^{AB}
 \operatorname{Tr}\big(\cA_{A}\,\delta \cA_{B}\big)\Big|_{\pi=\alpha}
=
\frac{1}{\langle \alpha \tilde{\alpha}\rangle \langle \tilde{\alpha} \beta\rangle^{2}}
\int_{\mathbb{E}^{4}} \mathrm{vol}_{4}\,\epsilon^{AB}
 \operatorname{Tr}\big(\cA_{A}\,\delta \cA_{B}\big)\Big|_{\pi=\tilde{\alpha}} .
\end{equation}
Let $\mu^{A}$ and $\hat\mu^{A}$ be a basis of the unprimed spinor space, normalised such that
$[\mu\hat\mu]=\epsilon_{AB}\mu^{A}\hat\mu^{B}=1$.
Since the space is two–dimensional, any spinor $X^{A}$ admits the expansion
$X^{A}=p\,\mu^{A}+q\,\hat\mu^{A}$.
Contracting with $\hat\mu_{A}$ and $\mu_{A}$ and using antisymmetry of the spinor
inner product $[XY]=\epsilon_{AB}X^{A}Y^{B}$ yields
$p=[X\hat\mu]$ and $q=-[X\mu]$, so that
\begin{equation}\label{Unit Spinor}
	X^{A}=[X\hat\mu]\mu^{A}-[X\mu]\hat\mu^{A}.
\end{equation}
Using \eqref{Unit Spinor} in the gauge-field component spinors $\cA_A$ and $\delta \cA_A$ (at fixed $\pi$) gives $\epsilon^{AB}\operatorname{Tr}(\cA_{A}\delta \cA_{B})
=\operatorname{Tr}([\cA\mu][\delta \cA\hat\mu]-[\cA\hat\mu][\delta \cA\mu])$,
which leads directly to 
\ie \label{6dboundeom}
\left.\frac{1}{\langle\alpha \beta\rangle^2} \operatorname{Tr}([\cA \mu][\delta \cA \hat{\mu}]-[\cA \hat{\mu}][\delta \cA \mu])\right|_{\pi=\alpha}=\left.\frac{1}{\langle {\tilde{\alpha}} \beta\rangle^2} \operatorname{Tr}(\cA \mu][\delta \cA\hat{\mu}]-[\cA \hat{\mu}][\delta \cA \mu])\right|_{\pi=\tilde{\alpha}}.
\fe

Now, the boundary condition we impose at the two simple poles of $\Omega$ is 
\begin{align}\label{boundary conditions}
&(\mathcal{O} - c)\,[\cA\mu]\Big|_{\pi=\alpha}
=
\sigma\,\frac{\langle \alpha \beta \rangle}{\langle \tilde{\alpha} \beta \rangle}\,
(\mathcal{O} + c)\,[\cA\mu]\Big|_{\pi=\tilde{\alpha}},\nonumber\\
    &(\mathcal{O} - c)\,[\cA\hat{\mu}] \Big|_{\pi=\alpha}
=
\sigma^{-1}\,
\frac{\langle \alpha \beta \rangle}{\langle \tilde{\alpha} \beta \rangle}\,
(\mathcal{O} + c)\,[\cA\hat{\mu}] \Big|_{\pi=\tilde{\alpha}},
\end{align}
 where we require the operator $\mathcal{O} \in \textrm{End }\mathfrak{g}$ to be skew-symmetric, i.e.
$\mathcal{O}^t = -\mathcal{O}$, and where $c$ is a complex constant. Consequently, the boundary equation of motion is satisfied using the boundary conditions \eqref{boundary conditions}, since
\ie 
&\frac{1}{\langle \alpha \beta \rangle^{2}}
\operatorname{Tr}\Bigg(
\sigma\,\frac{\langle \alpha \beta \rangle}{\langle \tilde{\alpha} \beta \rangle}
\frac{(\mathcal{O} + c)}{(\mathcal{O}-c)}[\cA\mu]\Big|_{\pi=\tilde{\alpha}}
\cdot
\sigma^{-1}\frac{\langle \alpha \beta \rangle}{\langle \tilde{\alpha} \beta \rangle}
\frac{(\mathcal{O} + c)}{(\mathcal{O}-c)}[\delta \cA\hat{\mu}]\Big|_{\pi=\tilde{\alpha}}
\nonumber\\&-\sigma^{-1}\frac{\langle \alpha \beta \rangle}{\langle \tilde{\alpha} \beta \rangle}
\frac{(\mathcal{O} + c)}{(\mathcal{O}-c)}[\cA\hat{\mu}]\Big|_{\pi=\tilde{\alpha}}\cdot
\sigma\frac{\langle \alpha \beta \rangle}{\langle \tilde{\alpha} \beta \rangle}
\frac{(\mathcal{O} + c)}{(\mathcal{O}-c)}[\delta \cA\mu]\Big|_{\pi=\tilde{\alpha}}
\Bigg)\\=&\left.\frac{1}{\langle {\tilde{\alpha}} \beta\rangle^2} \operatorname{Tr}(\cA \mu][\delta \cA\hat{\mu}]-[\cA \hat{\mu}][\delta \cA \mu])\right|_{\pi=\tilde{\alpha}}.
\fe 

Although the dependence on $\mathcal{O}$ in these boundary conditions may seem arbitrary, in later sections, we shall investigate the specialisation of the operator $\mathcal{O}$ to a skew-symmetric solution of the modified classical Yang--Baxter equation associated with $c=1$ or $i$, whereby the boundary conditions shall resemble those used to derive the 2d Yang--Baxter sigma model from 4d Chern--Simons theory in \cite{Delduc:2019whp}. As shown in \cite{Delduc:2019whp}, although the derivations of the 2d Yang--Baxter sigma model in the two cases of $c=1$ and $c=i$ are similar, they differ in a subtle way in that the field space of the models (prior to gauge fixing)
are different, and correspond to $G\times G$ for $c=1$ and $G_{\mathbb{C}}$ for $c=i$. 

In fact, we could be more general, and consider boundary conditions of the form 
\begin{align}\label{boundary conditionsP}
&P\,[\cA\mu]\Big|_{\pi=\alpha}
=
\sigma\,\frac{\langle \alpha \beta \rangle}{\langle \tilde{\alpha} \beta \rangle}\,
\,[\cA\mu]\Big|_{\pi=\tilde{\alpha}},\nonumber\\
    &P\,[\cA\hat{\mu}] \Big|_{\pi=\alpha}
=
\sigma^{-1}\,
\frac{\langle \alpha \beta \rangle}{\langle \tilde{\alpha} \beta \rangle}\,
\,[\cA\hat{\mu}] \Big|_{\pi=\tilde{\alpha}},
\end{align}
where $P\in \textrm{End }\mathfrak{g}$ satisfies $P^t=P^{-1}$, with the choice $P=\left(\frac{\mathcal{O}-c}{\mathcal{O}+c}\right)$ corresponding to the boundary conditions in \eqref{boundary conditions}. In what follows, although our focus will be on the boundary conditions \eqref{boundary conditions}, we shall express many of our results such that they depend on the operator $P=\left(\frac{\mathcal{O}-c}{\mathcal{O}+c}\right)$.

\subsection{Derivation of 4d Field Theory}
We shall now proceed to derive a four-dimensional field theory from the 6d Chern--Simons theory on twistor space with boundary conditions \eqref{boundary conditions}.
To perform the localisation analysis, it shall be convenient to rewrite the $(0,1)$-connection $\cA$ in terms of new variables $\cA'$ and a group-valued field $\hat{h}$ via
\begin{align}\label{gauge transformation}
    \cA = \hat{h}^{-1} \cA' \hat{h} + \hat{h}^{-1} \bar{\partial} \hat{h}.
\end{align}
This relation has the form of a gauge transformation, but is in fact a change of variables that separates the part of the connection which is pure gauge, particularly along the $\mathbb{CP}^1$ fibre. This part of the connection is pure gauge since any complex bundle on a
 $\mathbb{CP}^1$ fibre  that is topologically trivial is also generically holomorphically trivial.
%This parametrisation has redundancies, since different choices of $(\cA', \hat{h})$ can represent the same $\cA$, but these redundancies are useful as they allow one to impose convenient constraints on $\cA'$. 
Notably, the fibre component of $\cA'$ is equal to zero, so that the remaining field may be interpreted as a four-dimensional anti-self-dual Yang--Mills connection.

%The internal gauge redundancy of the parametrisation $\cA = \hat{h}^{-1} \cA' \hat{h} + \hat{h}^{-1} \bar{\partial} \hat{h}$ allows one to impose the condition $\cA'_0 = 0$, where $\cA'_0$ denotes the component of the $(0,1)$-connection along the $\mathbb{CP}^1$ fibre. Indeed, under the internal gauge transformations
%$\cA' \mapsto \hat{g}^{-1} \cA' \hat{g} + \hat{g}^{-1} \bar{\partial} \hat{g},
%$
%the fibre component transforms as
%$
%\cA'_0 \mapsto \hat{g}^{-1} \cA'_0 \hat{g} + \hat{g}^{-1} \bar{\partial}_0 \hat{g},
%$
%which is the standard transformation law for a connection in one complex dimension. Locally on $\mathbb{CP}^1$, this allows $\cA'_0$ to be gauged away by solving a first-order equation for $\hat{g}$. Imposing $\cA'_0 = 0$ removes only the auxiliary fibre component of the connection and leaves the spacetime components $\cA'_A$ unconstrained, apart from ordinary four-dimensional gauge transformations.

As a result, $\cA'$ may be regarded as a $(0,1)$-connection with support entirely along the $\mathbb{E}^4$ directions, which is precisely the structure required for its interpretation, via the Penrose--Ward correspondence, as an anti-self-dual Yang--Mills connection on $\mathbb{E}^4$.
After fixing $\cA'_0 = 0$, there remains a residual internal gauge freedom consisting of gauge transformations involving smooth functions that are independent of the $\mathbb{CP}^1$ fibre coordinate. We can fix this symmetry by fixing the value of the group-valued field $\hat{h}$ at one of the poles, which we choose to be $\pi = \beta$, by imposing $\hat{h}\rvert_{\beta} = \mathrm{id}$. 

However, because gauge transformations are restricted at the remaining poles by boundary conditions associated with the meromorphic structure of the theory, the values of $\hat{h}$ at $\pi = \alpha$ and $\pi = \tilde{\alpha}$,
\begin{align}
    h = \hat{h}\rvert_{\alpha},
\qquad
\tilde{h} = \hat{h}\rvert_{\tilde{\alpha}},
\end{align}
cannot be gauged away. These fields therefore become genuine dynamical degrees of freedom, representing edge modes localised at the poles. As a consequence, the holomorphic Chern--Simons action localises to an effective four-dimensional theory on $\mathbb{E}^4$ whose dynamics depend only on these edge-mode fields. Also, after imposing $\cA'_0 = 0$, $\cA'$ only has legs in a two-dimensional $(0,1)$ subspace, so any triple wedge product is forced to vanish. In terms of the redefined gauge field $\cA'$, the transformed action takes the form
\begin{align}\label{gauged action}
S_{\mathrm{hCS}_6}
=&
\frac{1}{2\pi i}
\int_{\mathbb{PT}}
\Omega \wedge \mathrm{Tr}\left(\cA' \wedge \bar{\partial} \cA'\right)
+
\frac{1}{2\pi i}
\int_{\mathbb{PT}}
\bar{\partial}\Omega \wedge \mathrm{Tr}\left(\cA' \wedge \bar{\partial}\hat{h}\,\hat{h}^{-1}\right)
\nonumber\\&-
\frac{1}{6\pi i}
\int_{\mathbb{PT}}
\Omega \wedge \mathrm{Tr}\left(
\hat{h}^{-1}\bar{\partial}\hat{h}
\wedge
\hat{h}^{-1}\bar{\partial}\hat{h}
\wedge
\hat{h}^{-1}\bar{\partial}\hat{h}
\right).
\end{align}
%The first term in the action is the bulk term, it will go away from the bulk equation of motion
%Because the internal redundancy acts on $A'$ as$A' \mapsto \hat{g}^{-1} A' \hat{g} + \hat{g}^{-1} \bar{\partial} \hat{g},$ the fibre component transforms as$A'_0 \mapsto \hat{g}^{-1} A'_0 \hat{g} + \hat{g}^{-1} \bar{\partial}_0 \hat{g}.$Hence one may set $A'_0 = 0$ by choosing $\hat{g}$ to solve$\bar{\partial}_0 \hat{g} = - A'_0 \hat{g},$ a first-order equation along the $\mathbb{CP}^1$ direction. Locally on the fibre this equation always admits solutions, so $A'_0$ is pure gauge with respect to the internal redundancy. The remaining components $A'_A$ are unconstrained (they transform covariantly under the residual $\mathbb{CP}^1$-independent internal symmetry), allowing $A'$ to be interpreted as the spacetime gauge field.
The latter two terms in the action \eqref{gauged action} localise to the poles of the meromorphic form $\Omega$, and the remaining obstruction to deriving the four-dimensional theory is the presence of a genuine bulk term and the residual dependence on the field $\cA'$. Both issues are resolved by invoking the bulk equations of motion for $\cA'$. Varying  the action, which is the only bulk contribution, yields the equation of motion
\begin{equation}\label{eom gauge field}
\bar\partial \cA' +\cA'\wedge \cA'= 0 .
\end{equation}
Imposing the gauge condition $\cA'_0 = 0$ and decomposing $\cA' = \cA'_A \,\bar e^A$, this equation implies
\begin{equation}\label{solution0}
\bar\partial_0 \cA'_A = 0 ,
\end{equation}
so that the components $\cA'_A$ are holomorphic along the $\mathbb{CP}^1$ fibre, and thus there are no nontrivial bulk contributions to the action  \eqref{gauged action}. 

Since $A'_A$ has homogeneous weight one under rescalings of the fibre coordinate, it follows that its $\mathbb{CP}^1$-dependence is necessarily linear, and may be written as
\begin{equation}
\cA'_A = \pi^{A'} A_{AA'} ,\label{solution A}
\end{equation}
where $A_{AA'}$ is independent of $\pi$.
We now recall the boundary condition at the double pole $\pi=\beta$, which is $\cA_A|_\beta = 0$. Recalling that $\hat h|_\beta = \mathrm{id}$, this condition translates directly into $\cA'_A|_\beta = 0$, which implies $\beta^{A'} A_{AA'} = 0$. Hence $A_{AA'}$ must be proportional to $\langle \pi\beta\rangle$, and we may write
\begin{equation}\label{solution}
\cA'_A = \langle \pi\beta\rangle\, B_A,
\end{equation}
for some $\pi$-independent field $B_A$. Substituting this result into the field redefinition
\begin{equation}
\cA = \hat h^{-1} \cA' \hat h + \hat h^{-1} \bar\partial \hat h,
\end{equation}
and extracting the spacetime component yields
\begin{equation}\label{gauge field component}
\cA_A = \langle \pi\beta\rangle\,\mathrm{Ad}_{\hat h}^{-1}(B_A)
+ \pi^{A'}\,\hat h^{-1}\partial_{AA'} \hat h.
\end{equation}
From \eqref{boundary conditions} and \eqref{gauge field component}, we find 
\begin{align}\label{boundary conditions2}
    &(\mathcal{O} - c)\Big[
\langle \alpha \beta \rangle \,
\mathrm{Ad}^{-1}_{h}(B_A)
+
\alpha^{A'} h^{-1} \partial_{A A'} h
\Big] \mu^{A}
\\=&
\sigma \,
\frac{\langle \alpha \beta \rangle}{\langle \tilde{\alpha} \beta \rangle}
(\mathcal{O} + c)\Big[
\langle \tilde{\alpha} \beta \rangle \,
\mathrm{Ad}^{-1}_{\tilde{h}}(B_A)
+
\tilde{\alpha}^{A'} \tilde{h}^{-1} \partial_{A A'} \tilde{h}
\Big] \mu^{A},\nonumber\\&
(\mathcal{O} - c)
\Big[
\langle \alpha \beta \rangle \,
\mathrm{Ad}^{-1}_{h}(B_A)
+
\alpha^{A'} h^{-1} \partial_{A A'} h
\Big]\hat{\mu}^{A}
\\=&
\sigma^{-1}
\frac{\langle \alpha \beta \rangle}{\langle \tilde{\alpha} \beta \rangle}
(\mathcal{O} + c)
\Big[
\langle \tilde{\alpha} \beta \rangle \,
\mathrm{Ad}^{-1}_{\tilde{h}}(B_A)
+
\tilde{\alpha}^{A'} \tilde{h}^{-1} \partial_{A A'} \tilde{h}
\Big]\,\hat{\mu}^{A}.
\end{align}
It follows that the solutions to these constraints can be expressed as
%satisfying the remaining boundary conditions can be expressed as
\begin{align}\label{b-solutions}
\mathrm{Ad}_{h^{-1}} (B_A) = \hat{b}\,\mu_A - b\,\hat{\mu}_A, \quad
\mathrm{Ad}_{\tilde{h}^{-1}} (B_A) = \tilde{\hat{b}}\,\mu_A - \tilde{b}\,\hat{\mu}_A.
\end{align}
We define $b = \mathrm{Ad}^{-1}_{h}[B{\mu}]$ and $\hat{b} = \mathrm{Ad}^{-1}_{h}[B{\hat{\mu}}]$, together with their tilded counterparts. By using \eqref{boundary conditions}, we find that the resulting fields are related by the following relations 
\begin{align}
\tilde{b} - \tilde{j}
= \sigma^{-1} \bigg(\frac{\mathcal{O}-c}{\mathcal{O}+c}\bigg)\,(b - j),\quad
\tilde{\hat{b}} - \tilde{\hat{j}}
= \sigma \bigg(\frac{\mathcal{O}-c}{\mathcal{O}+c}\bigg)\,(\hat{b} - \hat{j}),
\end{align}
and moreover from \eqref{b-solutions}, they satisfy $\tilde{b}=\Lambda b$ and $\tilde{\hat{b}}=\Lambda \hat{b}$, where $\Lambda = \mathrm{Ad}^{-1}_{\tilde{h}} \,\mathrm{Ad}_{h},
$ and we obtain  
\begin{align}\label{b expression1}
	b
	= U_{+}\left(
	Pj-\sigma\,\tilde{j}
	\right),\quad \hat{b}
	= U_{-}\left(
	P\hat{j}-\sigma^{-1}\hat{\tilde{j}}
	\right),
\end{align}
where, we define 
\begin{align}\label{Operators}
	U_{\pm}
	= \left(
	P - \sigma^{\pm 1}\Lambda
	\right)^{-1},\quad P=\left(\frac{\mathcal{O}-c}{\mathcal{O}+c}\right),
\end{align}
as well as the currents $j$, $\tilde{j}$, $\hat{j}$ and $\tilde{\hat{j}}$ as
\begin{align}\label{currents}
j
&=
\langle \alpha \beta \rangle^{-1}
\,\mu^{A}\alpha^{A'}\,
h^{-1}\partial_{A A'}h,
&
\hat{j}
&=
\langle \alpha \beta \rangle^{-1}
\,\hat{\mu}^{A}\alpha^{A'}\,
h^{-1}\partial_{A A'}h,
\nonumber\\[0.5em]
\tilde{j}
&=
\langle \tilde{\alpha} \beta \rangle^{-1}
\,\mu^{A}\tilde{\alpha}^{A'}\,
\tilde{h}^{-1}\partial_{A A'}\tilde{h},
&
\tilde{\hat{j}}
&=
\langle \tilde{\alpha} \beta \rangle^{-1}
\,\hat{\mu}^{A}\tilde{\alpha}^{A'}\,
\tilde{h}^{-1}\partial_{A A'}\tilde{h}.
\end{align}

The operators $U_\pm$ are subject to the following identities, which play an important role in the analysis
\begin{align}
&\Lambda^T U^T_{\pm}=-\sigma^{\mp 1} U_{\mp}P, 	\quad  U^T_{\pm}\Lambda^T=-\sigma^{\mp 1} P U_{\mp},\nonumber\\&
PU_{\pm}+U_{\mp}^T P^T=1, \quad U_\pm P +P^T U_\mp^T=1\label{Useful Identities}.
\end{align}

Coming back to \eqref{gauged action} since the first term vanishes upon imposing the bulk equations of motion, we are left with 
\begin{align}
S_{\mathrm{IFT}_4}
  &= \frac{1}{2\pi i}
    \int_{\mathbb{PT}} \bar\partial \Omega \wedge
    \operatorname{Tr}\big( \mathcal{A}' \wedge \bar\partial \hat h \hat h^{-1} \big)
    - \frac{1}{6\pi i}
\int_{\mathbb{PT}}
\Omega \wedge \mathrm{Tr}\left(
\hat{h}^{-1}\bar{\partial}\hat{h}
\wedge
\hat{h}^{-1}\bar{\partial}\hat{h}
\wedge
\hat{h}^{-1}\bar{\partial}\hat{h}
\right),\nonumber\\&=
\frac{K}{2\pi i}
\int
 \bar{e}_0\,
\bar{\partial_0}
\left(
\frac{1}{\langle \pi \alpha \rangle}
\right)
\frac{1}{\langle \pi \tilde{\alpha} \rangle \langle \pi \beta \rangle^{2}}
\,
\wedge e^{0} \wedge e^{A} \wedge e_{A}
\wedge \bar{e}^{C} \wedge \bar{e}^{D}\,
\mathrm{Tr}
\!\left(
A'_{C}\,\bar{\partial}_{D}\hat{h}\,\hat{h}^{-1}
\right)
\nonumber
\\[0.5em]
&\quad
+ \frac{K}{2\pi i}
\int
\bar{e}_0 \,
\bar{\partial_0}
\left(
\frac{1}{\langle \pi \tilde{\alpha} \rangle}
\right)
\frac{1}{\langle \pi \alpha \rangle \langle \pi \beta \rangle^{2}}\wedge
\,
e^{0} \wedge e^{A} \wedge e_{A}
\wedge \bar{e}^{C} \wedge \bar{e}^{D}\,
\mathrm{Tr}
\!\left(
A'_{C}\,\bar{\partial}_{D}\hat{h}\,\hat{h}^{-1}
\right)\nonumber\\&- \frac{1}{6\pi i}
\int_{\mathbb{PT}}
\Omega \wedge \mathrm{Tr}\left(
\hat{h}^{-1}\bar{\partial}\hat{h}
\wedge
\hat{h}^{-1}\bar{\partial}\hat{h}
\wedge
\hat{h}^{-1}\bar{\partial}\hat{h}
\right).
\end{align}
In evaluating the boundary variation of the action, we make use of the identities $e^{C} \wedge e_{C} \wedge \bar{e}^{A} \wedge \bar{e}^{B}
= -2\,\mathrm{vol}_{4}\,\epsilon^{AB},$ where
$\mathrm{vol}_{4}
=
\mathrm{d}x^{0} \wedge \mathrm{d}x^{1} \wedge \mathrm{d}x^{2} \wedge \mathrm{d}x^{3},$ together with $\frac{1}{2\pi i}
\int_{\mathbb{CP}^{1}}
e^{0} \wedge \bar{e}^{0}\,
\bar{\partial}_{0}
\left(
\frac{1}{\langle \pi \alpha \rangle}
\right)
f(\pi)
=
f(\alpha)$ and \eqref{solution}. The effective four-dimensional action takes the form 
\ie 
S_{\mathrm{IFT}_4}=&\frac{K}{\langle \alpha \tilde{\alpha} \rangle}
\int \frac{1}{\langle \alpha\beta \rangle^{2}}\,
\mathrm{vol}_{4}\,\epsilon^{CD}\,
\mathrm{Tr}\left(B_{C}\,\bar{\partial}_{D}h\,h^{-1}\right)\Big|_{\pi=\alpha}\\&-
\frac{K}{\langle \alpha \tilde{\alpha} \rangle}
\int \frac{1}{\langle \tilde{\alpha}\beta \rangle^{2}}
\mathrm{vol}_{4}\,\epsilon^{CD}\,
\mathrm{Tr}\left(B_{C}\,\bar{\partial}_{D}\tilde{h}\,\tilde{h}^{-1}\right)\Big|_{\pi=\tilde{\alpha}}\\&- \frac{1}{6\pi i}
\int_{\mathbb{PT}}
\Omega \wedge \mathrm{Tr}\!\left(
\hat{h}^{-1}\bar{\partial}\hat{h}
\wedge
\hat{h}^{-1}\bar{\partial}\hat{h}
\wedge
\hat{h}^{-1}\bar{\partial}\hat{h}
\right),\\=&\frac{K}{\langle \alpha \tilde{\alpha} \rangle}
\int \frac{1}{\langle \alpha\beta \rangle}\,
\mathrm{vol}_{4}\,\epsilon^{AB}\,
\mathrm{Tr}\left(\mathrm{Ad}_{h^{-1}} (B_A)\,h^{-1}\bar{\partial}_{B}h\right)\Big|_{\pi=\alpha}\\&-
\frac{K}{\langle \alpha \tilde{\alpha} \rangle}
\int \frac{1}{\langle \tilde{\alpha}\beta \rangle^{2}}
\mathrm{vol}_{4}\,\epsilon^{AB}\,
\mathrm{Tr}\left(\mathrm{Ad}_{h^{-1}} (B_A)\,\tilde{h}^{-1}\bar{\partial}_{B}\tilde{h}\right)\Big|_{\pi={\tilde{\alpha}}}\\&- \frac{1}{6\pi i}
\int_{\mathbb{PT}}
\Omega \wedge \mathrm{Tr}\!\left(
\hat{h}^{-1}\bar{\partial}\hat{h}
\wedge
\hat{h}^{-1}\bar{\partial}\hat{h}
\wedge
\hat{h}^{-1}\bar{\partial}\hat{h}
\right).\label{ift41}
\fe 
Now, using $\bar{\partial}_{A} = \pi^{A'}\partial_{A A'}
$ and \eqref{b-solutions}, we finally get 
\begin{align}
S_{\mathrm{IFT}_4}
&=
\frac{K}{\langle \alpha \tilde{\alpha} \rangle}
\int
\frac{1}{\langle \alpha \beta \rangle}\,
\mathrm{vol}_{4}\,\epsilon^{AB}\,
\mathrm{Tr}\left[
\big(\hat{b}\,\mu_{A} - b\,\hat{\mu}_{A}\big)\,
\alpha^{B'}h^{-1}\bar{\partial}_{B B'} h
\right]
\nonumber
\\[0.6em]
&
-
\frac{K}{\langle \alpha \tilde{\alpha} \rangle}
\int
\frac{1}{\langle \tilde{\alpha} \beta \rangle}\,
\mathrm{vol}_{4}\,\epsilon^{AB}\,
\mathrm{Tr}\!\left[
\big(\tilde{\hat{b}}\,\mu_{A} - \tilde{b}\,\hat{\mu}_{A}\big)\,
\tilde{\alpha}^{B'} \tilde{h}^{-1}\bar{\partial}_{B B'} \tilde{h}
\right]\nonumber\\&- \frac{1}{6\pi i}
\int_{\mathbb{PT}}
\Omega \wedge \mathrm{Tr}\!\left(
\hat{h}^{-1}\bar{\partial}\hat{h}
\wedge
\hat{h}^{-1}\bar{\partial}\hat{h}
\wedge
\hat{h}^{-1}\bar{\partial}\hat{h}
\right),\nonumber\\&= \frac{K}{\langle \alpha \tilde{\alpha} \rangle}
	\int_{\mathbb{E}^4} \mathrm{vol}_4\,
	\mathrm{Tr}\!\bigl( 
	b(\hat{j} - \Lambda^{T}\hat{\tilde{j}}) 
	- \hat{b}(j - \Lambda^{T}\tilde{j})
	\bigr)
	+ S_{\mathrm{WZ}_4},
\end{align}
where
\begin{align}
S_{\mathrm{WZ}_4}
= \frac{K}{\langle \alpha \tilde{\alpha} \rangle}
\int_{\mathbb{E}^4 \times [0,1]}
\mathrm{vol}_4 \wedge \mathrm{d}\rho \;
\mathrm{Tr}\!\left(
h^{-1}\partial_\rho h \cdot [j,\hat{j}]
- \tilde{h}^{-1}\partial_\rho \tilde{h} \cdot [\tilde{j},\hat{\tilde{j}}]
\right).
\end{align}
Here, the final term $- \frac{1}{6\pi i}
\int_{\mathbb{PT}}
\Omega \wedge \mathrm{Tr}\!\left(
\hat{h}^{-1}\bar{\partial}\hat{h}
\wedge
\hat{h}^{-1}\bar{\partial}\hat{h}
\wedge
\hat{h}^{-1}\bar{\partial}\hat{h}
\right)$ in \eqref{ift41}
has been identified with a four-dimensional Wess–Zumino term, denoted by $S_{\mathrm{WZ}_4}$. A detailed derivation of this term can be found in \cite{Cole:2024ess}. Importantly, this term is unaffected by the deformation and depends only on the fields $h$ and $\tilde{h}$. Finally, by using  \eqref{b expression1}, \eqref{Operators} and \eqref{Useful Identities}, we find a 4d field theory given by
\ie \label{ift4actionn}
	S_{\textrm{IFT}_4}=& \frac{K}{\langle \alpha \tilde{\alpha} \rangle}
	\int_{\mathbb{E}^4} \mathrm{vol}_4\,
	\mathrm{Tr}\bigl( 
	U_{+}(
	Pj-\sigma\,\tilde{j}
	)(\hat{j} - \Lambda^{T}\hat{\tilde{j}}) 
	-U_{-}(
	P\hat{j}-\sigma^{-1}\hat{\tilde{j}}
	)(j - \Lambda^{T}\tilde{j})
	\bigr)
	+ S_{\mathrm{WZ}_4},\\=&\frac{K}{\langle \alpha \tilde{\alpha} \rangle}
	\int_{\mathbb{E}^4} \mathrm{vol}_4\,
	\mathrm{Tr}\bigl(j(P^T U_+^{T}-U_- P)\hat{j}+\tilde{j}(U_+^T P^T-PU_-)\hat{\tilde{j}}-2\sigma\tilde{j}U_+^T\hat{j}+2\sigma^{-1} jU_-\hat{\tilde{j}}
	\bigr)\\
    &
	+ S_{\mathrm{WZ}_4}.
\fe

\subsection{Equations of Motion } 
Let us now describe the equations of motion of the 4d field theory we have derived with the aim of relating them to the anti-self-dual Yang--Mills equations. 
Consider an infinitesimal variation of the group element introduced through 
\( g \rightarrow g e^{\epsilon} \approx g(1+\epsilon) \), 
with \( \epsilon \) belonging to the Lie algebra \( \mathfrak{g} \). Hence, $g^{-1}\delta g =  \epsilon$ and varying \( j_{\mu} = g^{-1} \partial_{\mu} g \) gives $\delta j_\mu= \partial_{\mu} \epsilon + [ j_{\mu}, \epsilon ]$. Together with the identities listed in \eqref{Useful Identities}, the variation of the operator $U_{\pm}$ plays a crucial role. This leads to the following relations, which are used in the derivation of the equations of motion:
%\begin{align}\delta(U_+ X)&=U_+\delta( X)+\delta(U_+ )X\nonumber\\&=U_+\delta( X)+\sigma\delta \Lambda (U_+)\end{align}
\begin{align}
&\delta(U_{\pm}X) = U_{\pm}(\delta X) + U_{\pm}\left[\, \tilde{h}^{-1}\delta\tilde{h},\, U_{\mp}^{T}P^{T}(X) \right]
- P^{T}U_{\mp}^{T}\left[\, h^{-1}\delta h,\, U_{\pm}(X) \right].
\end{align}
The variation of the 4d field theory action is then
\begin{align}\delta_{h} S_{4}
&=
\frac{K}{\langle \alpha \tilde{\alpha} \rangle}
\int_{\mathbb{E}^{4}} \mathrm{vol}_{4}\;
\mathrm{Tr}\Bigg(\delta j\hat{j}-2\delta j(U_-P\hat{j}-\sigma^{-1}U_-\hat{\tilde{j}})-j\delta \hat{j}+2(U_+ Pj-U_+\tilde{j})\delta\hat{j}\nonumber\\&+2\epsilon[U_-P\hat{j},U_+Pj]+2\epsilon[U_-\hat{\tilde{j}},U_+\tilde{j}]-2\epsilon\sigma[U_-P \hat{j},U_+\tilde{j}]-2\epsilon\sigma^{-1}[U_-\hat{\tilde{j}},U_+Pj]+\epsilon[j,\hat{j}]
\Bigg)\nonumber\\&	= \frac{K}{\langle \alpha \tilde{\alpha} \rangle}
	\int_{\mathbb{E}^{4}} \mathrm{vol}_{4}\;
	\mathrm{Tr}\big(\delta j \hat{j}-2\delta j(\hat{b})-j\delta\hat{j}+2(b)\delta\hat{j}+\epsilon[{\hat {b},b}]+\epsilon[j,\hat{j}]\big).
\end{align}
For the variation with respect to 
$h$, the corresponding equation of motion takes the form
\begin{align}\label{EOMh}
-\frac{\mu^{A}\alpha^{A'}}{\langle \alpha\beta\rangle}\,
\partial_{AA'} \hat{b}
&+\frac{\hat{\mu}^{A}\alpha^{A'}}{\langle \alpha\beta\rangle}\,
\partial_{AA'} b
+[\hat{j}, b] - [j, \hat{b}] - [\hat{b}, b] = 0,
\end{align}
A similar variation with respect to $\tilde{h}$
 yields the equation
\begin{align}\label{EOMtildeh}-\frac{\mu^{A}\tilde{\alpha}^{A'}}{\langle \tilde{\alpha}\beta\rangle}\,
\partial_{AA'} \hat{\tilde{b}}
&+\frac{\hat{\mu}^{A}\tilde{\alpha}^{A'}}{\langle \tilde{\alpha}\beta\rangle}\,
\partial_{AA'} \tilde{b}
+[\hat{\tilde{j}}, \tilde{b}] - [\tilde{j}, \hat{\tilde{b}}] - [\hat{\tilde{b}}, \tilde{b}] = 0 .
\end{align}

\subsection{4d ASDYM and Lax formulation}\label{asdymlax}
We now turn to the details of the integrability of the 4d field theory. This can be established by showing that the resulting equations of motion \eqref{EOMh} and \eqref{EOMtildeh} are equivalent to the four-dimensional anti-self-dual Yang–Mills equations. To this end, it suffices to rewrite the equations of motion in terms of the solution for $B^{A}$.

We first evaluate $\mathrm{Ad}_{h^{-1}}[B_A, B^A]$ in the $h$-frame. Using the fact that the adjoint action is an automorphism, together with \eqref{b-solutions} and \eqref{adjoint commutator}, we obtain
\begin{align}\label{commutator part}
\mathrm{Ad}_{h^{-1}}\left[ B_A, B^A \right]
&=
\left[
\mathrm{Ad}_{h^{-1}} (B_A),\,
\mathrm{Ad}_{h^{-1}} (B^A)
\right]\nonumber\\&
=
\big[
\hat{b}\,\mu_A - b\,\hat{\mu}_A,\,
\hat{b}\,\mu^A - b\,\hat{\mu}^A
\big]\nonumber\\&=
\mu_{A}\mu^{A}\,[\hat{b},\hat{b}]
-
\mu_{A}\hat{\mu}^{A}\,[\hat{b},b]
-
\hat{\mu}_{A}\mu^{A}\,[b,\hat{b}]
+
\hat{\mu}_{A}\hat{\mu}^{A}\,[b,b]\nonumber\\&=2 [\hat{b}, b]. 
\end{align}
Here, we use the fact that the spinors $\mu_A$ and $\hat{\mu}_A$ satisfy 
$
\mu_A \mu^{A} = 0,
\hat{\mu}_A \hat{\mu}^{A} = 0,
\mu_A \hat{\mu}^{A} = -1$ and
 $\hat{\mu}_A \mu^{A} = 1$. Next, we compute $\mathrm{Ad}_{h^{-1}}\left(
\alpha^{A'} \partial_{A A'} B^{A}
\right)$. This calculation is carried out in detail in Appendix~\ref{ap B}, and yields
\begin{align}\label{p2}
\mathrm{Ad}_{h^{-1}}\left(
\alpha^{A'} \partial_{A A'} B^{A}
\right)
&=\mu^{A} \alpha^{A'} \partial_{A A'} \hat{b}
-
\hat{\mu}^{A} \alpha^{A'} \partial_{A A'} b+{\langle \alpha\beta\rangle}
\big[ j , \hat{b} \big]
-{\langle \alpha\beta\rangle}
 \big[ \hat{j} ,\, b \big].
\end{align}
Combining the contributions \eqref{commutator part} and \eqref{p2}, we obtain an equation of motion equivalent to \eqref{EOMh}, expressed in terms of $B^{A}$
\begin{align}\alpha^{A'} \partial_{AA'} B^A
+ \frac{1}{2}\,\langle \alpha \beta \rangle
\bigl[ B_A, B^A \bigr]
= 0.\label{eom simplify1}
\end{align}
Similarly for $\tilde{h}$,\begin{align}\label{eom simplify2}
    \tilde{\alpha}^{A'} \partial_{AA'} B^A
+ \frac{1}{2}\,\langle \tilde{\alpha}\,\beta \rangle
\bigl[ B_A, B^A \bigr]
= 0.
\end{align}
These equations of motions can be recast as ASDYM equations. From \eqref{eom gauge field}, the ``$0A$" component of the curvature,
$\cF'_{0A} = \bar{\partial}_{0} \cA'_{A} - \bar{\partial}_{A} \cA'_{0} + [\cA'_{0}, \cA'_{A}]$,
simplifies upon imposing the gauge condition $\cA'_{0}=0$, leading to the constraint
$\bar{\partial}_{0} \cA'_{A}=0$. The remaining $\cF'_{AB}$ component of \eqref{eom gauge field}
then yields the equation
$\bar{\partial}_{A} \cA'_{B} - \bar{\partial}_{B} \cA'_{A} + [\cA'_{A}, \cA'_{B}] = 0$,
which expresses the integrability condition for the gauge field in the $(0,1)$ directions. Recall that, by definition of the frame, $\bar{\partial}_{A}
=
\pi^{A'} \partial_{A A'}$ and using \eqref{solution A}, we obtain $\bar{D}'_{A}
=
\bar{\partial}_{A} + \cA'_{A}
=
\pi^{A'} \partial_{A A'} + \pi^{A'} A_{A A'}
=
\pi^{A'} \big( \partial_{A A'} + A_{A A'} \big).$ We define the spacetime covariant derivative
$
D_{A A'} = \partial_{A A'} + A_{A A'}.
$ Then we can write 
\begin{equation}\bar{D}'_{A} = \pi^{A'} D_{A A'}.
\end{equation}
This is the key relation linking the twistor connection to a connection on $\mathbb{E}^{4}$. Now we turn to the second hCS equation $\cF'_{AB} = 0$. By definition, $\cF'_{AB} = [\bar{D}'_A, \bar{D}'_B]$, and using the equation of motion, we get
\begin{align}\label{eom hs6}
\cF'_{AB}=[\pi^{A'} D_{AA'}, \pi^{B'} D_{BB'}]
=
\pi^{A'} \pi^{B'} [D_{AA'}, D_{BB'}]=\pi^{A'} \pi^{B'} \cF_{AA'BB'} =0, 
\end{align}
where
$$[D_{AA'}, D_{BB'}]=\cF_{AA'BB'}
=
\partial_{AA'} A_{BB'}
-
\partial_{BB'} A_{AA'}
+
\left[ A_{AA'}, A_{BB'} \right].$$
Next, we show \eqref{eom hs6} is precisely the ASDYM condition. For that we use the spinor decomposition of the field strength. Any real two-form $\cF_{\mu\nu}$ in four dimensions can be written in spinor
notation as $\cF_{AA'BB'}
=
\cF_{\mu\nu}\,\sigma^\mu_{AA'} \sigma^\nu_{BB'} .$ Representation theory of $\mathrm{SL}(2,\mathbb{C}) \times \mathrm{SL}(2,\mathbb{C})$
tells us that any such $\cF_{AA'BB'}$ satisfies
\begin{equation}
\cF_{AA'BB'}
=
\epsilon_{AB}\,\Phi_{A'B'}
+
\epsilon_{A'B'}\,\Psi_{AB} ,
\end{equation}
with
\begin{equation}
\Phi_{A'B'} = \Phi_{(A'B')},
\qquad
\Psi_{AB} = \Psi_{(AB)} .
\end{equation}
The symmetric spinor $\Phi_{A'B'}$ encodes the self-dual part $\cF^{(+)}$, and
$\Psi_{AB}$ encodes the anti-self-dual part $\cF^{(-)}$. Now compute
\begin{align}
\pi^{A'} \pi^{B'} \cF_{AA'BB'}
&=
\pi^{A'} \pi^{B'}
\bigl(
\epsilon_{AB}\,\Phi_{A'B'}
+
\epsilon_{A'B'}\,\Psi_{AB}
\bigr)
\nonumber \\
&=
\epsilon_{AB}\,\pi^{A'} \pi^{B'} \Phi_{A'B'}
+
\pi^{A'} \pi^{B'} \epsilon_{A'B'} \Psi_{AB} .
\end{align}
Since $\pi^{A'} \pi^{B'}$ is symmetric and $\epsilon_{A'B'}$ is antisymmetric,
we have $\pi^{A'} \pi^{B'} \epsilon_{A'B'} = 0 .$ So, we get $\pi^{A'} \pi^{B'} \cF_{AA'BB'}
=
\epsilon_{AB}\,\pi^{A'} \pi^{B'} \Phi_{A'B'} .$ Hence the twistor equation \eqref{eom hs6} becomes
\begin{equation}
\epsilon_{AB}\,\pi^{A'} \pi^{B'} \Phi_{A'B'} = 0
\qquad
\forall\, \pi^{A'} .
\end{equation}
Since the factor $\epsilon_{AB}$ is nonzero, we must have $\pi^{A'} \pi^{B'} \Phi_{A'B'} = 0,
\forall\, \pi^{A'} .$
But $\Phi_{A'B'}$ is symmetric, and we may regard it as a $2\times 2$ symmetric matrix.
 We define $P(\pi) := \pi^{A'} \pi^{B'} \Phi_{A'B'} $, a homogeneous quadratic polynomial in $\pi$. If $P(\pi) = 0$ for all
$\pi^{A'}$, every coefficient must vanish, so
\begin{equation}
\Phi_{A'B'} = 0 .
\end{equation}
As recalled earlier, $\Phi_{A'B'}$ is exactly the self-dual part of $\cF$. Thus
\begin{equation}
\Phi_{A'B'} = 0
\qquad \Longleftrightarrow \qquad
\cF^{(+)} = 0 .
\end{equation}
This is precisely the anti-self-dual Yang--Mills (ASDYM) equation:
\begin{equation}
\cF = \cF^{(-)}
\qquad
\text{(purely anti-self-dual curvature)} .
\end{equation}
Thus we have that
\begin{align}
   \cF'_{AB} = \pi^{A'} \pi^{B'} \cF_{AA'BB'} = 0
\;\;\Longleftrightarrow\;\;
\text{ASD Yang--Mills on } \mathbb{E}^4 \label{ASYDM}.
\end{align}

From the twistor discussion, the ASDYM is equivalent to \eqref{ASYDM}. Also, from \eqref{solution A},  we have $A_{AA'} = \beta_{A'} {B}_A$. We assume, $\beta_{A'}$ is a fixed spinor (constant in spacetime), $B_A(x)$ is Lie-algebra valued. So
$A_{BB'} = \beta_{B'} B_B$,
 and $A_{AA'} = \beta_{A'} B_A.$ Using \eqref{solution A} and \eqref{ASYDM}, the ASYDM simplifies to
\begin{align}\label{AYSDM2}
    \langle \pi \beta \rangle\, \pi^{A'} \partial_{AA'} B_B
- \langle \pi \beta \rangle\, \pi^{B'} \partial_{BB'} B_A
+ \langle \pi \beta \rangle^{2} \bigl[ B_A, B_B \bigr]
= 0,
\end{align}
where $\pi^{A'}$ is the homogeneous coordinate on the twistor
$\mathbb{CP}^1$. This condition must hold for every $\pi^{A'} \in \mathbb{CP}^1$. Pick a fixed spinor basis $\alpha^{A'}$, $\tilde{\alpha}^{A'}$ with
$\langle \alpha \tilde{\alpha} \rangle \neq 0$. Any $\pi^{A'}$ can be written uniquely as
\begin{equation}
\pi^{A'} = c_1 \alpha^{A'} + c_2 \tilde{\alpha}^{A'} ,
\end{equation}
for some complex coefficients $c_1, c_2$. To express $c_1, c_2$ in invariant form, contract with $\alpha_{A'}$ and
$\tilde{\alpha}_{A'}$:
\begin{align}
\langle \pi \alpha \rangle
&= \pi^{A'} \alpha_{A'}
= c_1 \langle \alpha \alpha \rangle
+ c_2 \langle \tilde{\alpha} \alpha \rangle
= c_2 \langle \tilde{\alpha} \alpha \rangle
= - c_2 \langle \alpha \tilde{\alpha} \rangle , \\\nonumber
\langle \pi \tilde{\alpha} \rangle
&= \pi^{A'} \tilde{\alpha}_{A'}
= c_1 \langle \alpha \tilde{\alpha} \rangle
+ c_2 \langle \tilde{\alpha} \tilde{\alpha} \rangle
= c_1 \langle \alpha \tilde{\alpha} \rangle .
\end{align}
Hence
\begin{equation}
c_1 = \frac{\langle \pi \tilde{\alpha} \rangle}{\langle \alpha \tilde{\alpha} \rangle},
\qquad
c_2 = - \frac{\langle \pi \alpha \rangle}{\langle \alpha \tilde{\alpha} \rangle} .
\end{equation}
 So
\begin{equation}\label{pi2}
\pi^{A'}
=
\frac{1}{\langle \alpha \tilde{\alpha} \rangle}
\left(
\langle \pi \tilde{\alpha} \rangle \, \alpha^{A'}
-
\langle \pi \alpha \rangle \, \tilde{\alpha}^{A'}
\right) .
\end{equation}
Contract \eqref{pi2} with $\beta_{A'}$

\begin{align}
\langle \pi \beta \rangle
= \pi^{A'} \beta_{A'} =
\frac{
\langle \pi \tilde{\alpha} \rangle \langle \alpha \beta \rangle
-
\langle \pi \alpha \rangle \langle \tilde{\alpha} \beta \rangle
}{
\langle \alpha \tilde{\alpha} \rangle
} .\label{pibeta}
\end{align}
It is convenient to define the antisymmetric derivative combination $K_{AB A'}
=
\partial_{AA'} B_B
-
\partial_{BA'} B_A .$ 
Using \eqref{pi2}, we get
\begin{equation}
\pi^{A'} K_{AB A'}
=
\frac{1}{\langle \alpha \tilde{\alpha} \rangle}
\left(
\langle \pi \tilde{\alpha} \rangle \, \alpha^{A'} K_{AB A'}
-
\langle \pi \alpha \rangle \, \tilde{\alpha}^{A'} K_{AB A'}
\right) .\label{k}
\end{equation}
Using, \eqref{pi2},\eqref{pibeta} and\eqref{k}, we get
\begin{align}
0
&=
\langle \pi \beta \rangle
\left(
\pi^{A'} K_{AB A'}
+
\langle \pi \beta \rangle [B_A, B_B]
\right)
\nonumber \\
&=
\frac{
\langle \pi \tilde{\alpha} \rangle \langle \alpha \beta \rangle
-
\langle \pi \alpha \rangle \langle \tilde{\alpha} \beta \rangle
}{
\langle \alpha \tilde{\alpha} \rangle
}
\nonumber \\
&\quad \times
\Biggl\{
\frac{1}{\langle \alpha \tilde{\alpha} \rangle}
\left(
\langle \pi \tilde{\alpha} \rangle \,
\alpha^{A'} K_{AB A'}
-
\langle \pi \alpha \rangle \,
\tilde{\alpha}^{A'} K_{AB A'}
\right)
\nonumber \\
&\qquad\qquad
+
\frac{
\langle \pi \tilde{\alpha} \rangle \langle \alpha \beta \rangle
-
\langle \pi \alpha \rangle \langle \tilde{\alpha} \beta \rangle
}{
\langle \alpha \tilde{\alpha} \rangle
}
[B_A, B_B]
\Biggr\} .
\end{align}
We can pull out the common constant factor
$1/\langle \alpha \tilde{\alpha} \rangle^{2}$.
The remaining expression is a polynomial in
$\langle \pi \tilde{\alpha} \rangle$ and
$\langle \pi \alpha \rangle$. So the coefficient of $\langle \pi \beta \rangle \langle \pi \tilde{\alpha} \rangle$
is proportional to
\begin{equation}
\alpha^{A'} K_{AB A'}
+
\langle \alpha \beta \rangle [B_A, B_B]
=
0,
\end{equation}
and the coefficient of $\langle \pi \beta \rangle \langle \pi \alpha \rangle$
gives
\begin{equation}
\tilde{\alpha}^{A'} K_{AB A'}
+
\langle \tilde{\alpha} \beta \rangle [B_A, B_B]
=
0 .
\end{equation} These are the same as the spacetime equations of motion of the 4d field theory in \eqref{eom simplify1} and \eqref{eom simplify2}. 

%\textbf{Lax Formulation}:

The ASDYM equations are integrable, with a Lax pair given by
\begin{align}
L^{(B)} &= \langle \pi \hat{\gamma} \rangle^{-1}\, \hat{\mu}^{A}\, \pi^{A'} D_{AA'} , 
\qquad
M^{(B)} = \langle \pi \gamma \rangle^{-1}\, \mu^{A}\, \pi^{A'} D_{AA'} ,
\end{align}
following  \cite{Cole:2023umd}. Here, convenient normalisations have been chosen that involve constant spinors $\gamma$ and $\hat{\gamma}$ that will appear when we consider symmetry reductions in subsequent sections. 
These operators satisfy the zero-curvature (flatness) condition
\begin{equation}
\bigl[ L^{(B)}, M^{(B)} \bigr] = 0
\end{equation}
for all $\pi^{A'} \in \mathbb{CP}^1$, which is equivalent to the
anti-self-dual Yang--Mills equations. We shall thus refer to the 4d field theory we have derived as $\textrm{IFT}_4$ in what follows.

%%%%%%%%%%%%%%%%%%%%%%%%%%%%%%%%%%%%%%%%%%%%%%%%%%%%%%%%%%%%%%%%%%%%%%%%%%%%%%%%%
\section{Symmetry reduction of 6d CS to 4d CS}\label{6dCSto4dCS}
%\textcolor{blue}{MA : Go back to general $c$ in boundary condition, change notation to match earlier section, also move this section earlier, before semi-local symmetry discussion  }

In this section, we shall describe how the 6d holomorphic Chern--Simons theory on twistor space that we considered in Section \ref{Sec2} can be symmetry reduced to 4d Chern--Simons theory with disorder surface defects. The symmetry reduction of the action essentially follows from \cite{Bittleston:2020hfv} and \cite{Cole:2023umd}; what is new here is the novel set of boundary conditions for the 6d twistor Chern--Simons theory.

The symmetry reduction relies on our freedom to add any $(1,0)$-form to the 6d gauge field, i.e., we may perform the shift 
\ie 
\mathcal{A} \rightarrow  \rho_0 e^0 +\rho_A e^A + \mathcal{A}
\fe 
without changing the action \eqref{6daction}. A judicious choice of $\rho_A$ results in a four-dimensional gauge field, that can be identified with the gauge field of 4d Chern--Simons theory.  

To perform the reduction, we impose invariance of the six-dimensional gauge field with respect to certain vector fields. To this end, we shall pick a unit norm spinor $\gamma_{A'}$, that selects a complex structure on $\mathbb{E}^4\subset \mathbb{PT}$ corresponding to the point $\pi_{A'}=\gamma_{A'}$. The spinor $\mu^A$ then allows us to define the following basis of one-forms in this complex structure : 
\begin{equation}\label{dual v}
\begin{aligned}
\mathrm{d} z & =\mu_A \gamma_{A^{\prime}} \mathrm{d} x^{A A^{\prime}}, & \mathrm{d} \bar{z} & =\hat{\mu}_A \hat{\gamma}_{A^{\prime}} \mathrm{d} x^{A A^{\prime}}, \\
\mathrm{d} w & =-\hat{\mu}_A \gamma_{A^{\prime}} \mathrm{d} x^{A A^{\prime}}, & \mathrm{d} \bar{w} & =\mu_A \hat{\gamma}_{A^{\prime}} \mathrm{d} x^{A A}.
\end{aligned}
\end{equation}
Given a coframe $\{\textrm{d} z, \textrm{d}\bar z, \textrm{d}w, \textrm{d}\bar w\}$, its dual frame
$\{\partial_z, \partial_{\bar z}, \partial_w, \partial_{\bar w}\}$ is defined by
\begin{equation}
\langle\partial_j,\theta^i\rangle = \delta_j{}^i ,
\end{equation}
i.e.
\begin{equation}
 \langle\partial_w,\textrm{d} w\rangle = 1,
\qquad
 \langle\partial_w,\textrm{d} z\rangle = 0,
\qquad
 \langle\partial_w,\textrm{d} \bar z\rangle = 0,
\qquad
 \langle\partial_w,\textrm{d} \bar{w}\rangle = 0,
\end{equation}
and similarly for the other directions.

The vector fields dual to $\textrm{d}z$ and $\textrm{d}\overline{z}$ are then 
\ie 
\chi= \hat{\mu}^A \hat{\gamma}^{A^{\prime}} \partial_{A A^{\prime}}=\partial_z, \quad \bar{\chi}=\mu^A \gamma^{A^{\prime}} \partial_{A A^{\prime}} =\partial_{\bar{z}},
\fe 
and symmetry reduction shall be performed with respect to these vector fields. The remaining dual vector fields are 
\ie 
\kappa=\mu^A \hat{\gamma}^{A^{\prime}} \partial_{A A^{\prime}}=-\partial_w, \quad \bar{\kappa}=\hat{\mu}^A \gamma^{A^{\prime}} \partial_{A A^{\prime}}=\partial_{\bar{w}}.
\fe

The reduction is performed by demanding
\ie 
\mathcal{L}_\chi{\mathcal{A}}=\mathcal{L}_{\bar{\chi}}{\mathcal{A}}=0,
\fe 
The shifted gauge field also satisfies $\iota_\chi{\mathcal{A}}=\iota_{\bar{\chi}}{\mathcal{A}}=0$, meaning it has no $\textrm{d}z$ or $\textrm{d}\bar{z}$ legs, and the explicit form of the 4d gauge field is 
\ie 
{A}=\bar{e}^0 {\mathcal{A}}_0+\left(\iota_\kappa {\mathcal{A}}-\frac{\langle\pi \hat{\gamma}\rangle}{\langle\pi \gamma\rangle} \iota_{\bar{\chi}} {\mathcal{A}}\right) \mathrm{d} w+\left(\iota_{\bar{\kappa}} {\mathcal{A}}+\frac{\langle\pi \gamma\rangle}{\langle\pi \hat{\gamma}\rangle} \iota_\chi {\mathcal{A}}\right) \mathrm{d} \bar{w}.
\fe 

The 6d to 4d reduction is then performed by contracting the bivector $\chi \wedge \bar{\chi}$ with the Lagrangian density of the twistor space Chern-Simons theory. 
Using
\ie 
\iota_{\chi \wedge \bar{\chi}} \Omega=
-K \frac{\langle\pi \gamma\rangle\langle\pi \hat{\gamma}\rangle}{\langle\pi \alpha\rangle\langle\pi \tilde{\alpha}\rangle\langle\pi \beta\rangle^2} e^0=\omega,
\fe 
it can be shown that 
\ie 
\iota_{\chi \wedge \bar{\chi}}(\Omega \wedge \operatorname{CS}({\mathcal{A}}))=\omega \wedge \operatorname{CS}({A}), \label{redlag}
\fe 
and the symmetry reduction results in the 4d Chern--Simons theory action 
\begin{equation}
\frac{1}{2 \pi {i}} \int_{\Sigma \times \mathbb{C P}^1} \omega \wedge \operatorname{Tr}\left({{A}} \wedge \mathrm{d} {{A}}+\frac{2}{3} {{A}} \wedge {{A}} \wedge {{A}}\right),
\end{equation}
where, crucially, 
\ie \label{matchcon}
{{A}}_w&=-\frac{[\mathcal{A} \mu]}{\langle\pi \gamma\rangle}\\ {{A}}_{\bar{w}}&=-\frac{[\mathcal{A} \hat{\mu}]}{\langle\pi \hat{\gamma}\rangle}
\fe
have appropriate singularities at $\gamma$ and $\hat{\gamma}$. In other words, the symmetry reduction has generated disorder surface defects at the locations of zeroes of $\omega$, despite the fact that such defects are not present in the 6d CS theory. 

Now, let us turn to the reduction of 6d Chern--Simons boundary conditions to 4d Chern--Simons boundary conditions. 
Recall that the 6d Chern--Simons boundary conditions are 
\begin{align}\label{boundary conditions2b}
&(\mathcal{O} - c)\,[\cA\mu]\Big|_{\pi=\alpha}
=
\sigma\,\frac{\langle \alpha \beta \rangle}{\langle \tilde{\alpha} \beta \rangle}\,
(\mathcal{O} + c)\,[\cA\mu]\Big|_{\pi=\tilde{\alpha}}\nonumber\\
    &(\mathcal{O} - c)\,[\cA\hat{\mu}] \Big|_{\pi=\alpha}
=
\sigma^{-1}\,
\frac{\langle \alpha \beta \rangle}{\langle \tilde{\alpha} \beta \rangle}\,
(\mathcal{O} + c)\,[\cA\hat{\mu}] \Big|_{\pi=\tilde{\alpha}},
\end{align}
and $\cA|_{\pi=\beta}=0$. Using \eqref{matchcon}, we immediately find the 4d Chern--Simons boundary conditions 
\begin{equation}
(\mathcal{O} - c)\left.{{A}}_w\right|_{\pi=\alpha}=(\mathcal{O} + c)\left.t s {{A}}_w\right|_{\pi=\tilde{\alpha}},\left.\quad (\mathcal{O} - c){{A}}_{\bar{w}}\right|_{\pi=\alpha}=(\mathcal{O} + c)\left.t^{-1} s {{A}}_{\bar{w}}\right|_{\pi=\tilde{\alpha}},
\end{equation}
and $A|_{\pi=\beta}=0$, where we have defined (following the conventions of \cite{Cole:2023umd})
\begin{equation}
	r_{+}
	=
	K\,\frac{\langle \alpha\gamma\rangle \langle \alpha\hat{\gamma}\rangle}
	{\langle \alpha\tilde{\alpha}\rangle \langle \alpha\beta\rangle^{2}},
	\qquad
	r_{-}
	=
	- K\,\frac{\langle \tilde{\alpha}\gamma\rangle \langle \tilde{\alpha}\hat{\gamma}\rangle}
	{\langle \alpha\tilde{\alpha}\rangle \langle \tilde{\alpha}\beta\rangle^{2}},
\end{equation}

\begin{equation}\label{seqn}
	s
	=
	\sqrt{-\frac{r_{-}}{r_{+}}}
	=
	\frac{\langle \alpha\beta\rangle}{\langle \tilde{\alpha}\beta\rangle}
	\sqrt{
		\frac{\langle \tilde{\alpha}\gamma\rangle \langle \tilde{\alpha}\hat{\gamma}\rangle}
		{\langle \alpha\gamma\rangle \langle \alpha\hat{\gamma}\rangle}
	},
\end{equation}

\begin{equation}\label{teqn}
	t
	=
	\sigma\, s\,
	\frac{\langle \tilde{\alpha}\beta\rangle \langle \alpha\hat{\gamma}\rangle}
	{\langle \alpha\beta\rangle \langle \tilde{\alpha}\hat{\gamma}\rangle}.
\end{equation}

It shall be convenient to specify an inhomogeneous coordinate on $\mathbb{CP}^1$, that is, $\zeta =\pi_{2'}/\pi_{1'}$ on the patch where $\pi_{1'}\neq 0$. Moreover, we shall specify the the remaining spinors in terms of the inhomogeneous coordinates, that is, we pick 
\begin{equation}
\alpha_{A^{\prime}}=\left(1, \alpha_{+}\right), \quad \tilde{\alpha}_{A^{\prime}}=\left(1, \alpha_{-}\right), \quad \beta_{A^{\prime}}=(0,1),
\end{equation}
implying 
\begin{equation}
\langle\tilde{\alpha} \beta\rangle=\langle\alpha \beta\rangle=1, \quad\langle\tilde{\alpha} \alpha\rangle=\alpha_{+}-\alpha_{-}.
\label{abchoices}
\end{equation}

Now, we shall further specify $\sigma$ in order to make a connection a more familiar integrable field theory. That is, we shall pick 
\ie \label{sigmachoice}
\sigma = \frac{\langle \alpha \gamma  \rangle }{\langle \tilde{\alpha} \gamma \rangle }= \frac{\langle \tilde{\alpha} \hat{\gamma}  \rangle }{\langle {\alpha} \hat{\gamma} \rangle },
\fe 
whereby $t=1$ and $s=1$.
Moreover, we shall pick $\mathcal{O}$ to be a solution, $R\in \textrm{End }\mathfrak{g} $, of the modified classical Yang--Baxter equation 
given by
\begin{equation}
[R \mathsf{x}, R \mathsf{y}]-R([R \mathsf{x}, \mathsf{y}]+[\mathsf{x}, R \mathsf{y}])=-c^2[\mathsf{x}, \mathsf{y}],
\end{equation}
for $c= i$ or $c=1$, and $\mathsf{x},\mathsf{y}\in \mathfrak{g}$. The 4d Chern--Simons boundary conditions then become 
\begin{equation}
(R - c)\left.{{A}}_w\right|_{\zeta=\alpha_+}=(R + c)\left. {{A}}_w\right|_{\zeta=\alpha_-},\left.\quad (R - c){{A}}_{\bar{w}}\right|_{\zeta=\alpha_+}=(R + c)\left.{{A}}_{\bar{w}}\right|_{\zeta=\alpha_-},
\label{4dBC}
\end{equation}
and $A|_{\pi=\beta}=0$ that are the familiar boundary conditions that realise the Yang--Baxter sigma model, as shown in \cite{Delduc:2019whp}, as long as we identify $\alpha_+$ with $-c\eta$ and $\alpha_-$ with $c\eta$ in \cite{Delduc:2019whp}.  The choice \eqref{sigmachoice} is a sensible one, since, if we pick $\gamma$ to correspond to $\zeta =1$ and $\hat{\gamma}$ to correspond to $\zeta=-1$, the second equality in \eqref{sigmachoice}  just corresponds to 
\ie 
\frac{-c\eta -1}{c\eta -1} = \frac{c\eta +1}{-c\eta +1}.
\fe 

In the following, we shall explore the consequences of the restrictions \eqref{sigmachoice} and $\mathcal{O}=R$.

\section{Modified Classical Yang--Baxter Equation, Semi-local Symmetry and a 4d Yang--Baxter Sigma Model}\label{semilcs}

Let us now investigate the implications of restricting $\sigma$ as in \eqref{sigmachoice} and picking $\mathcal{O}=R$, i.e., a solution of the modified classical Yang--Baxter equation, for the $\textrm{IFT}_4$

 Recall that the boundary conditions \eqref{4dBC} used to derive the two-dimensional YB sigma model exhibits gauge invariance, and fixing this gauge invariance leads us to an $\textrm{IFT}_2$ with a single field \cite{Delduc:2019whp}.\footnote{If we did not fix this gauge symmetry in the boundary conditions, the two-dimensional IFT would take the form of a two-field model that admits a nontrivial gauge symmetry.} The gauge transformation of the boundary condition is 
\ie \label{gaugetx}
(R-c)( u^{-1}\partial_i u |_{\zeta=\alpha_+}+u^{-1}A_i u|_{\zeta=\alpha_+})=(R+c)\left(u^{-1}\partial_i u |_{\zeta=\alpha_-}+u^{-1}A_i  u |_{\zeta=\alpha_-} \right),
\fe 
where $i=w,\bar{w}$.
We can show that the boundary condition is gauge invariant when $(u|_{\alpha_+},u|_{\alpha_-})=(r,\tilde{r})$, where $(r,\tilde{r})\in G_R$, the Lie group generated by the Lie algebra
\begin{equation}
\mathfrak{g}_R:=\{((R-c) x,(R+c) x) \mid {x} \in \mathfrak{g}\}.
\end{equation}
Gauge invariance follows since
\ie (R-c)r^{-1}\partial_i r   =(R+c)\tilde{r} ^{-1}\partial_i \tilde{r},  \fe 
and since \ie 
(R-c)( r^{-1}A_i|_{\zeta=\alpha_+} r )=(R+c)\left(\tilde{r}^{-1}A_i|_{\zeta=\alpha_-}  \tilde{r} \right),
\fe 
is equivalent to the original boundary condition. To show the latter,
we need to express the boundary restrictions of the gauge fields as $(A_i|_{\alpha_+},A_i|_{\alpha_-})=((R+c)\tilde{x}_i,(R-c)\tilde{x}_i )$ for $\tilde{x}_i\in \mathfrak{g}$, and express $(r,\tilde{r})$ as exponentials of the Lie algebra elements $((R+c)y,(R-c)y )$ for $y\in \mathfrak{g}$.
%, implying that $(r^{-1}A_i|_{\zeta=\alpha_+} r,\tilde{r}^{-1}A_i|_{\zeta=\alpha_-} \tilde{r})\in ((R+c)z,(R-c)z )$ for some $z\in \mathfrak{g}$. 
The original boundary condition can then be retrieved using the identity $\textrm{Ad}_{\textrm{exp } X} = \textrm{exp} (\textrm{ad }X)$ for $X\in \mathfrak{g}$ and the modified classical Yang--Baxter equation, stated as 
\begin{equation}
[({R} \pm c) X,({R} \pm c) Y]=({R} \pm c)([X, {R} Y]+[{R} X, Y]), \quad X, Y \in \mathfrak{g},
\end{equation}
which implies that $\mathfrak{h}_{\pm}:=\textrm{im}(R\pm c)$ are Lie algebras.

%\textcolor{blue}{MA : Double check this}

%This also means that $(R+c)\partial_iu u^{-1}|_{x_+}=(R-c)\partial_iu u^{-1}|_{x_-}$, which is the remaining condition necessary to prove gauge invariance. 

In the case of 6d CS on twistor space, the restrictions \eqref{sigmachoice} and $\mathcal{O}=R$ lead us not to a gauge symmetry of the boundary conditions and the $\textrm{IFT}_4$, but to a \textit{semi-local symmetry}. We shall refer to the $\textrm{IFT}_4$ with these restrictions as $\textrm{IFT}^{\textrm{YB}}_4$, or the 4d Yang--Baxter sigma model. Indeed,  using \eqref{abchoices}, the boundary conditions with these restrictions are 
%This is because the formal gauge transformed boundary condition here is 
\begin{align}\label{boundary conditionsspecialized}
&(R - c)\,[\cA\mu]\Big|_{\pi=\alpha}
=
\frac{\langle \alpha \gamma  \rangle }{\langle \tilde{\alpha} \gamma \rangle }\,\,
(R + c)\,[\cA\mu]\Big|_{\pi=\tilde{\alpha}}\nonumber\\
    &(R- c)\,[\cA\hat{\mu}] \Big|_{\pi=\alpha}
=
\frac{\langle \alpha \hat{\gamma}  \rangle }{\langle \tilde{\alpha} \hat{\gamma} \rangle }\,
(R + c)\,[\cA\hat{\mu}] \Big|_{\pi=\tilde{\alpha}}.
\end{align}
These boundary conditions can also be stated as 
\ie \label{nobrd}
[\cA\mu]\Big|_{\pi=\alpha} = \langle {\alpha}\gamma \rangle (R+c)x\\
[\cA\mu]\Big|_{\pi=\tilde{\alpha}} =\langle \tilde{\alpha}\gamma \rangle  (R-c)x\\
[\cA\hat{\mu}]\Big|_{\pi=\alpha} = \langle {\alpha}\hat{\gamma} \rangle (R+c)y\\
[\cA\hat{\mu}]\Big|_{\pi=\tilde{\alpha}} =\langle \tilde{\alpha}\hat{\gamma} \rangle  (R-c)y
\fe 
where $x,y\in \mathfrak{g}$.

If we perform a gauge transformation with respect to a gauge parameter $u$ which satisfies $(u|_{\alpha_+},u|_{\alpha_-})\in G_R$, then we find that the first boundary condition in \eqref{boundary conditionsspecialized} is transformed to (using $[A\mu]=-[\mu A]$)
\begin{equation}
    (R-c)\mu^A( \alpha^{A'} u^{-1}\partial_{AA'}u  +u^{-1} {\mathcal{A}}_Au )|_{\pi=\alpha}=   \frac{\langle \alpha\gamma \rangle}{\langle \tilde{\alpha}\gamma \rangle} (R+c)\mu^A( \tilde{\alpha}^{A'}u^{-1}\partial_{AA'}u  +u^{-1}\mathcal{A}_Au )|_{\pi=\tilde{\alpha}}.
\end{equation}
Using 
\ie \label{symmetry reduction spinor}
X_{A^{\prime}}=\langle X \hat{\gamma}\rangle \gamma_{A^{\prime}}-\langle X \gamma\rangle \hat{\gamma}_{A^{\prime}}
\fe 
 we then find that the derivative terms in this  expression can be rewritten as
\ie 
& (R-c)\left(-\mu^A\langle \alpha \gamma\rangle \hat{\gamma}^{A^{\prime}} u^{-1}\partial_{AA'}  u +\mu^A\langle \alpha\hat{\gamma}\rangle \gamma^{A^{\prime}} u^{-1}\partial_{A A^{\prime}} u \right)|_{\pi=\alpha} \\
& =(R+c ) \frac{\langle\alpha \gamma\rangle}{\langle \tilde{\alpha}\gamma\rangle}\left(-\mu^A\langle\tilde{\alpha} \gamma\rangle \hat{\gamma}^{A^{\prime}} u^{-1}\partial_{A A^{\prime}} u   +\mu^A\langle\tilde{\alpha}\hat{\gamma}\rangle {\gamma}^{A^{\prime}} u^{-1} \partial_{A A^{\prime}} u   \right)|_{\pi=\tilde{\alpha}},
\fe
while the remaining terms are equivalent to the original boundary condition using \eqref{nobrd}. 
 Invariance of the boundary condition then follows if 
\ie 
\mu^A \gamma^{A^{\prime}} \partial_{A A^{\prime}} u=0,
\fe 
which implies that the gauge invariance is only semi-local.
Analogously, the remaining boundary condition is invariant if 
\ie 
\hat{\mu}^A \hat{\gamma}^{A^{\prime}} \partial_{A A^{\prime}} u=0.
\fe 

These semi-local symmetries of the boundary conditions should be manifest in the $\textrm{IFT}^{\textrm{YB}}_4$.
Semi-local symmetries are common in the study of 4d IFTs, and occur even in the 4d WZW model \cite{Nair:1990aa, Nair:1991ab}.
As explained in \cite{Cole:2023umd}, there is also a semi-local symmetry that arises from  transformations that preserve the conditions $\mathcal{A}|_{\beta}=0$. The conserved current associated with this symmetry determines the $B$-Lax operator that we constructed in Section \ref{asdymlax}. It is expected that the semi-local symmetry described in this section determines another Lax operator, called the $C$-Lax operator in \cite{Cole:2023umd}. 

In the preceding discussion, we have shown that $\mathcal{O}=R$ is a sufficient condition for a semi-local symmetry in the $\textrm{IFT}_4$. It is in fact also a necessary condition, as we explain in Appendix \ref{subsec:general_O_gauge_boundary}. 

\section{Symmetry Reduction of $\textrm{IFT}_4$ to $\textrm{IFT}_2$ }

In this section, we shall perform the symmetry reduction of the $\textrm{IFT}_4$ defined by \eqref{ift4actionn} to a two-dimensional field theory, that corresponds to the 2d Yang--Baxter sigma model when we specialise to $\textrm{IFT}_4^{\textrm{YB}}$.
The symmetry reduction is performed along the vector
fields dual to $\textrm{d}z$ and $\textrm{d}\bar z$. Under this reduction, $w$ and $\bar w$ become the coordinates of the
two-dimensional worldsheet $\Sigma$. The symmetry reduction imposes the condition
\begin{align}
    \partial_z h = \partial_{\bar z} h = 0,
\qquad
\partial_z \tilde h = \partial_{\bar z} \tilde h = 0.
\end{align}
Let us start by writing the IFT$_4$ action in terms of the currents \eqref{currents}. In two
dimensions, we set
\begin{equation}
\partial_+ \equiv \partial_w,
\qquad
\partial_- \equiv \partial_{\bar w} ,
\end{equation}
Starting from the definition 
$j
=
\langle \alpha \beta \rangle^{-1}
\, \mu^{A} \alpha^{A'} \,
h^{-1} \partial_{AA'} h$, we use the spinor expansion \eqref{symmetry reduction spinor} with $X = \alpha$
\begin{equation}
\alpha^{A'}
=
\langle \alpha \hat{\gamma} \rangle \, \gamma^{A'}
-
\langle \alpha \gamma \rangle \, \hat{\gamma}^{A'} .
\end{equation}
Therefore, the derivative term $h^{-1} \partial_{AA'} h$  appearing in $j$ becomes
\begin{equation}
\mu^{A} \alpha^{A'} \, \partial_{AA'}
=
\langle \alpha \hat{\gamma} \rangle \,
\mu^{A} \gamma^{A'} \, \partial_{AA'}
-
\langle \alpha \gamma \rangle \,
\mu^{A} \hat{\gamma}^{A'} \, \partial_{AA'} .
\end{equation}
Now identify the first term using \eqref{dual v} which is $\mu^{A} \gamma^{A'} \, \partial_{AA'}
=
\partial_{\bar z}.$ However, symmetry reduction imposes
$\partial_z h=\partial_z \tilde{h} = 0 ,$ so the first term drops out on reduced fields. The second term is precisely the derivative along $w$ (i.e.\ the surviving
worldsheet direction), since $w$ and $\bar w$ are the coordinates that
remain after reducing along $z$ and $\bar z$. Thus, on reduced fields, $\mu^{A} \alpha^{A'} \, \partial_{AA'}
\;\longrightarrow\;
\langle \alpha \gamma \rangle \, \partial_w
\;\equiv\;
\langle \alpha \gamma \rangle \, \partial_+ $. Substituting this back into $j$, we find
\begin{equation}
j
\;\longrightarrow\;
\langle \alpha \beta \rangle^{-1}
\langle \alpha \gamma \rangle \,
h^{-1} \partial_+ h
=
\frac{\langle \alpha \gamma \rangle}{\langle \alpha \beta \rangle} \,
J_+ .
\end{equation}
Similarly, replace
$
    \hat{j}
\;\longrightarrow\;
\frac{\langle \alpha \hat{\gamma} \rangle}{\langle \alpha \beta \rangle}
J_-,\quad  \tilde{j}
\;\longrightarrow\;
\frac{\langle \tilde{\alpha} \gamma \rangle}{\langle \tilde{\alpha} \beta \rangle}
\, \tilde{J}_+$ and $\tilde{\hat{j}}
\;\longrightarrow\;
\frac{\langle \tilde{\alpha} \hat{\gamma} \rangle}{\langle \tilde{\alpha} \beta \rangle}
\, \tilde{J}_-$, where, 
$$
J_+ = h^{-1} \partial_+ h,
\qquad
J_- = h^{-1} \partial_- h,
\qquad
\tilde{J}_+ = \tilde{h}^{-1} \partial_+ \tilde{h},
\qquad
\tilde{J}_- = \tilde{h}^{-1} \partial_- \tilde{h}.
$$
Then, the resulting 2-dimensional action is given by
\ie\label{ift22}
	S_{\text{IFT}_4} \;\rightsquigarrow\; S_{\text{IFT}_2}
	=&
	\int_{\Sigma} \mathrm{vol}_2 \,
	\mathrm{Tr}\Big(
	r_{+} J_{+}\big(P^TU_{+}^{T}-U_{-}P\big)J_{-}
	- r_{-} \tilde{J}_{+}\big(U_{+}^{T}P^T-PU_{-}\big)\tilde{J}_{-}
	\\&  + r_{+}\,\mathcal{L}_{\mathrm{WZ}}(h)
	+ r_{-}\,\mathcal{L}_{\mathrm{WZ}}(\tilde{h})	\\&- 2t\,\sqrt{-r_{+} r_{-}}\,
	\tilde{J}_{+} U_{+}^{T} J_{-}
	+ 2t^{-1}\,\sqrt{-r_{+} r_{-}}\,
	J_{+} U_{-} \tilde{J}_{-}
	\Big).
\fe
The symmetry reduction along $\partial_z$ and $\partial_{\bar z}$ maps the 4-form
$\mathrm{vol}_4
=
\frac{1}{12}\,
\varepsilon_{AB}\,\varepsilon_{CD}\,\varepsilon_{A'C'}\,\varepsilon_{B'D'}\,
\textrm{d}x^{AA'} \wedge \textrm{d}x^{BB'} \wedge \textrm{d}x^{CC'} \wedge \textrm{d}x^{DD'}$ to the 2-form $\mathrm{vol}_2$, defined by $\mathrm{vol}_2
= \iota_{\partial_z}\,\iota_{\partial_{\bar z}}\,\mathrm{vol}_4
= \textrm{d}\bar w \wedge \textrm{d}w
= \textrm{d}\sigma^{-} \wedge \textrm{d}\sigma^{+}.$
\subsection{Reduction to Yang--Baxter Sigma Model}\label{ybreduc}

We shall now specialise the $\textrm{IFT}_2$ in \eqref{ift22} to the case where
\ie 
\mathcal{O}=R,
\fe 
\ie
\sigma
=
\frac{1+\eta c}{1-\eta c}.
\fe 
This would correspond to the symmetry reduction of $\textrm{IFT}_4^{\textrm{YB}}$, that admits a semi-local symmetry associated with $R$. 
In this case 
\ie \label{coefficients value}
r_+&=-r_-\\
s&=1\\
t&=1.
\fe
The resulting 2d field theory is expected to enjoy a gauge symmetry corresponding to the group $G_R$, due to the gauge invariance of the 4d CS boundary condition \eqref{gaugetx}. As explained in Appendix \ref{subsec:general_O_gauge_boundary}, $\mathcal{O}=R$ is a necessary and sufficient condition for gauge symmetry of the 2d field theory defined by \eqref{ift22}.
This gauge symmetry can be fixed using $h=\tilde{h}$, whereby this two-field model ought to reduce to the Yang--Baxter sigma model. The reason that $h=\tilde{h}$ is the correct gauge fixing is because the field space is $D=G\times G$ or $G_{\mathbb{C}}$ when $c$ is equal to 1 or i, respectively. Since there is a $G_R$ gauge symmetry, modding out the corresponding redundancies in field space gives $G_R\backslash D$ or $G_R\backslash G_{\mathbb{C}}$. As explained, e.g., in \cite{Delduc:2019whp}, $G_R\backslash D$ can be parametrised by the \textit{diagonal} subgroup $G^\delta$, which corresponds to subgroup of $D$ determined by $h=\tilde{h}$. Analogously, $G_R\backslash G_{\mathbb{C}}$ can be parametrised by $G$ with elements determined by $h=\tilde{h}$.
%, and the relationship between $h$ and $\tilde{h}$ can be understood .

Let us show that this two-field model reduces to the Yang--Baxter sigma model when $h=\tilde{h}$, whereby $\Lambda|_{h=\tilde{h}}=1.$
Since $\Lambda = 1$,
\begin{equation}
U_{\pm} = \bigl(P - \sigma^{\pm 1}\bigr)^{-1}.
\end{equation}
Also, because $R^{T} = -R$, we have
\begin{equation}
P^{T} = P^{-1}.
\end{equation}
Therefore, we have
\begin{equation}
U_{\pm}^{T}
=
\bigl(P^{T} - \sigma^{\pm 1}\bigr)^{-1}
=
\bigl(P^{-1} - \sigma^{\pm 1}\bigr)^{-1}.
\end{equation}
Making the identification $\tilde{J}_{\pm} = J_{\pm}$, the action becomes
\begin{align}\label{lag}
S_{\text{IFT}_2}
	=&
	\int_{\Sigma} \mathrm{vol}_2 \,
\Tr\Big(
 r_{+}\, J_{+}\big(P^{T} U_{+}^{T} - U_{-} P\big) J_{-}
- r_{-}\, J_{+}\big(U_{+}^{T} P^{T} - P U_{-}\big) J_{-}
\nonumber\\
& - 2t \sqrt{-r_{+} r_{-}}\; J_{+} U_{+}^{T} J_{-}
+ 2t^{-1} \sqrt{-r_{+} r_{-}}\; J_{+} U_{-} J_{-}+(r_{+} + r_{-})\,\mathcal{L}_{\text{WZ}}(h)
\Big).
\end{align}
Using the identities in \eqref{Useful Identities} and setting $\Lambda^{T} = 1$ (since $\Lambda = 1$), we obtain
\begin{equation}
U_{+}^{T} = -\sigma^{-1} U_{-} P,
\qquad
U_{-}^{T} = -\sigma U_{+} P.
\end{equation}
Define $\mathcal{O}_1 = P^{T} U_{+}^{T} - U_{-} P$ and 
$\mathcal{O}_2 = U_{+}^{T} P^{T} - P U_{-}$ in \eqref{lag}. Simplifying $\mathcal{O}_1$ using $U_{+}^{T} = -\sigma^{-1} U_{-} P$, we obtain
\begin{align}
\mathcal{O}_1
&= P^{T}(-\sigma^{-1} U_{-} P) - U_{-} P \nonumber\\
&= -(\sigma^{-1}  P^{T} + 1)\, U_{-} P.
\end{align}
Now using $P^{T} = P^{-1}$ and $P$, $U_{+}$ and $U_-$ commute with each other, which simplifies $\mathcal{O}_1$ as
\begin{equation}
\mathcal{O}_1 = -(\sigma^{-1}  + P)\, U_{-}.
\end{equation}
Similarly, to simplify $\mathcal{O}_2$, we use $U_{+}^{T} = -\sigma^{-1} U_{-} P$, to obtain 
\begin{align}
\mathcal{O}_2= -(\sigma^{-1}  + P)\, U_{-}.
\end{align}
Therefore
\begin{equation}\label{AB}
\mathcal{O}_1 = \mathcal{O}_2 =-(\sigma^{-1}  + P)\, U_{-}
\qquad \text{when } \Lambda = 1.
\end{equation}
Rewriting the action using \eqref{AB}, it becomes
\begin{align}
S_{\text{IFT}_2}
	=&
	\int_{\Sigma} \mathrm{vol}_2 \,
\Tr\Big(
-(r_{+}-r_{-})\, J_{+} (\sigma^{-1} + P) U_{-} J_{-}
- 2t \sqrt{-r_{+} r_{-}}\, J_{+} U_{+}^{T} J_{-}
\nonumber\\&+ 2t^{-1} \sqrt{-r_{+} r_{-}}\, J_{+} U_{-} J_{-}+(r_{+} + r_{-})\,\mathcal{L}_{\text{WZ}}(h)
\Big).
\end{align}
We can use  $U_{+}^{T} = -\sigma^{-1} U_{-} P$ again to find 
\begin{align}
    -2 \sqrt{-r_{+} r_{-}}\, J_{+} U_{+}^{T} J_{-}=2\sigma^{-1}  \sqrt{-r_{+} r_{-}}\, J_{+} U_{-} P J_{-}.
\end{align}
Thus,
\begin{equation}
S_{\text{IFT}_2}
	=
	\int_{\Sigma} \mathrm{vol}_2 \,
\Tr\bigl(
J_{+} U_{-}\, \mathcal{M}(P)\, J_{-}
\bigr)+S_{WZ},
\end{equation}
where the remaining operator coefficient is
\begin{equation}
\mathcal{M}(P)
=
-(r_{+}-r_{-})(\sigma^{-1} + P)
+ 2t\sigma^{-1}  \sqrt{-r_{+} r_{-}}\, P
+ 2t^{-1} \sqrt{-r_{+} r_{-}}\,.
\end{equation}
Imposing \eqref{coefficients value}, which eliminates the Wess–Zumino part of the action and $\mathcal{M}(P)$ simplifies to
\begin{align}\label{YB Operator}
\mathcal{M}(P)
&=
r_{+}\Big(
- 2(\sigma^{-1} + P)
+ 2\sigma^{-1} \, P
+ 2 
\Big).
\end{align}
The operator sandwiched between $J_+$ and $J_-$ is $U_{-}(\mathcal{M}(P))$ is given by $U_{-} = (P - \sigma^{-1})^{-1}=-\sigma (1 - \sigma P)^{-1}.$ Therefore,
\begin{equation}\label{lkin}
S_{\text{IFT}_2}
	=
	2 (\sigma-1) r_{+}\int_{\Sigma} \mathrm{vol}_2 \,
\,
\Tr\!\Big(
J_{+} \big((1 - \sigma P)^{-1}\big)
\big(
  1-P
\big)
J_{-}
\Big).
\end{equation}
A identity can be derived for the YB-operator that can relate  $R$ and $P$, that is
\begin{align}\label{YB Operator1}
    (1 - \sigma P)^{-1}(1 - P)
&=
\frac{2}{1 + \sigma}\,(1 - \eta R)^{-1}
\qquad
\eta = \frac{\sigma - 1}{(\sigma + 1)c},
\nonumber\\&
=
(1 - \eta c)(1 - \eta R)^{-1},
\qquad
\sigma = \frac{1 + \eta c}{1 - \eta c}.
\end{align}
This identity is derived in Appendix \ref{ap C}.
From \eqref{YB Operator1}, the standard Yang--Baxter operator $(1-\eta R)^{-1}$ is expressed in terms of the $P$-operator, that is, it is proportional to $(1 - \sigma P)^{-1}(1 - P)$. Also, from \eqref{lkin}, we see that the the action is also proportional to $(1 - \sigma P)^{-1}(1 - P)$. As a result, the action can be written as
\begin{align}
S_{\text{IFT}_2}
	=
	C_1\int_{\Sigma} \mathrm{vol}_2 \Tr\Big(J_+\frac{1}{1-\eta R}J_-\Big)
\end{align}
where
$$C_1=4r_+\frac{(\sigma-1)}{(\sigma+1)}=4r_+c\eta.$$
This is precisely the action for the Yang--Baxter sigma model. Given that we have obtained the Yang--Baxter sigma model as a symmetry reduction of $\textrm{IFT}_4^{\textrm{YB}}$, and the equations of motion of the latter can be identified with the ASDYM equations, this means that we have obtained an embedding of the equations of motion of the Yang--Baxter sigma model in the ASDYM equations!

\section{2d IFT from 4d CS}

%\subsection{YB-sigma Model before Gauge-fixing}

In this section, we shall derive the two-dimensional field theory given in \eqref{ift4actionn} from the 4d Chern--Simons theory setup that was obtained via symmetry reduction in Section \ref{6dCSto4dCS}. 
%We shall then discuss the integrability of this theory in terms of the Lax connections it admits when the parent $\textrm{IFT}_4$ enjoys a semi-local symmetry.
We begin by parametrising the gauge field $\hat A$ by 
\begin{align}
    \hat A_{\bar\zeta}
=
\hat h^{-1}\partial_{\bar\zeta}\hat h,
\qquad
\hat A_I
=
\hat h^{-1} \mathscr{L}_I \hat h
+
\hat h^{-1}\partial_I \hat h,
\qquad
I = w, \bar w 
\end{align}
which corresponds to a formal gauge choice in which the $\bar\zeta$-component of the connection is pure gauge. In this parametrisation the fields $ \mathscr{L}_I$ turn out to define a meromorphic Lax connection on $\Sigma$, while residual gauge transformations act as ordinary two-dimensional gauge transformations. With this setup, the 4d CS action becomes
\begin{align}
    S_{\mathrm{CS4}}
=
-\,\frac{1}{2\pi i}
\int_{\Sigma \times \mathbb{CP}^1}
\textrm{d}\omega \wedge
\mathrm{Tr}\!\left(
\hat J \wedge \hat h^{-1} \mathscr{L} \hat h
\right)
\;+\; \textrm{WZ terms}.
\end{align}
Since $\omega$ is a meromorphic one-form on $\mathbb{CP}^1$, given explicitly by $\omega
=
-K\,
\frac{\langle \pi \gamma \rangle \langle \pi \hat{\gamma} \rangle}
{\langle \pi \alpha \rangle \langle \pi \hat{\alpha} \rangle \langle \pi \beta \rangle^{2}}
\,e^0$, it follows, in the distributional sense, that $\textrm{d}\omega
=
2\pi i
\sum_{p \in \mathrm{poles}(\omega)}
\mathrm{res}_p(\omega)\,\delta^{(2)}_p$. Using this identity, the action reduces to 
\begin{align}
S_{\mathrm{CS4}}
&= -\,\frac{1}{2\pi i}
\int_{\Sigma}
\int_{\mathbb{CP}^1}
\left(
2\pi i
\sum_{p}
\mathrm{Res}_p(\omega)\,\delta^{(2)}(p)
\right)
\wedge
\mathrm{Tr}\!\left(
\hat J \wedge \hat h^{-1} \mathscr{L} \hat h
\right) \\
&= - \sum_{p}
\mathrm{Res}_p(\omega)
\int_{\Sigma}
\mathrm{Tr}\!\left(
\hat J \wedge \hat h^{-1} \mathscr{L} \hat h
\right)\Big|_{p}.
\end{align}
The resulting localised $2$d action is given by
\begin{align}
    S_{2d}
=
r_+ \int_{\Sigma}
\mathrm{Tr}\!\left(
\hat J \wedge \hat h^{-1} \mathscr{L} \hat h
\right)\Big|_{\alpha}
+
r_- \int_{\Sigma}
\mathrm{Tr}\!\left(
\hat J \wedge \hat h^{-1} \mathscr{L} \hat h
\right)\Big|_{\bar\alpha}
\;+\; \text{WZ terms},
\end{align}
where 
\begin{align}
r_{+}
=
K\,
\frac{\langle \alpha \gamma \rangle \langle \alpha \hat{\gamma} \rangle}
{\langle \alpha \hat{\alpha} \rangle \langle \alpha \beta \rangle^{2}},
\qquad
\text{and}
\qquad
r_{-}
=
-\,K\,
\frac{\langle \tilde{\alpha} \gamma \rangle \langle \tilde{\alpha} \hat{\gamma} \rangle}
{\langle \alpha \hat{\alpha} \rangle \langle \tilde{\alpha} \beta \rangle^{2}}.
\end{align}
Note that we set
\ie 
\hat{h}=\textrm{id}
\fe 
in order to fix the residual gauge symmetry arising from transformations involving smooth functions on $\Sigma$. In what follows, we shall also define 
$$\hat h\big|_{\alpha} = h,\textrm{ }
\hat h\big|_{\bar\alpha} = \tilde h,\textrm{ }
\hat J\big|_{\alpha} = J, \textrm{ and } \hat J\big|_{\bar\alpha} = \tilde J .$$

To complete the construction, one must fix the meromorphic dependence of the Lax connection $ \mathscr{L}$ on the twistor coordinate $\pi\in\mathbb{CP}^1$ in a manner consistent with the meromorphic one-form $\omega$ and the boundary conditions at its poles. Since $\omega$ has zeros at $\pi\sim\gamma$ and $\pi\sim\hat\gamma$, regularity of the four-dimensional Chern--Simons action requires that $ \mathscr{L}$ may have at most simple poles at these points. The most general well-defined ansatz with this property is therefore
\begin{align}
    \mathscr{L}_w
=
\frac{\langle \pi \beta \rangle}{\langle \pi \gamma \rangle}\, M_w
+ N_w,
\qquad
 \mathscr{L}_{\bar w}
=
\frac{\langle \pi \beta \rangle}{\langle \pi \hat{\gamma} \rangle}\, M_{\bar w}
+ N_{\bar w},
\end{align}
where $ M_I$ and $ N_I$ are $\mathfrak g$-valued fields on $\Sigma$ independent of $\pi$.
The boundary conditions $A|_{\beta}=0$ and  
%\begin{align}
 %   \hat h\big|_{\beta} = \mathrm{id},
%\qquad
% \mathscr{L}\big|_{\beta} = 0,
%\end{align}
\begin{equation}
(\mathcal{O} - c)\left.{{A}}_w\right|_{\pi=\alpha}=(\mathcal{O} + c)\left.t s {{A}}_w\right|_{\pi=\tilde{\alpha}},\left.\quad (\mathcal{O} - c){{A}}_{\bar{w}}\right|_{\pi=\alpha}=(\mathcal{O} + c)\left.t^{-1} s {{A}}_{\bar{w}}\right|_{\pi=\tilde{\alpha}}.
\end{equation}
then uniquely determine the allowed Lax connection.
Solving for $M_w$, $M_{\bar w}$, $N_w$, and $N_{\bar w}$, we obtain
\begin{align}\label{lax1}
\frac{\langle \alpha \beta \rangle}{\langle \alpha \gamma \rangle}\mathrm{Ad}_h^{-1}(M_w)
&=
\left[
P - \sigma\, \mathrm{Ad}_{\tilde h}^{-1}\mathrm{Ad}_h
\right]^{-1}
\left(
t\, s \tilde J_w - P J_w
\right),
\qquad
\mathcal{N}_w = 0,
\nonumber\\[6pt]
\frac{\langle \alpha \beta \rangle}{\langle \alpha \hat{\gamma} \rangle}\mathrm{Ad}_h^{-1}(M_{\bar w})
&=
\left[
P - \sigma^{-1}\, \mathrm{Ad}_{\tilde h}^{-1}\mathrm{Ad}_h
\right]^{-1}
\left(
t^{-1} s \tilde J_{\bar w} - PJ_{\bar w}
\right),
\qquad
\mathcal{N}_{\bar w} = 0.
\end{align}
We also find it convenient to write these expressions in  the alternative forms
\begin{align}\label{lax2}
\frac{\langle \tilde{\alpha} \beta \rangle}{\langle \tilde{\alpha} \gamma \rangle}\mathrm{Ad}^{-1}_{\tilde{h}}(M_w)
&=
\left[
P^T - \sigma^{-1}\, \mathrm{Ad}_h^{-1}\mathrm{Ad}_{\tilde h}
\right]^{-1}
\left(
t^{-1}\, s^{-1} J_w - P^T \tilde J_w
\right),
\qquad
\mathcal{N}_w = 0,
\nonumber\\[6pt]
\frac{\langle \tilde{\alpha} \beta \rangle}{\langle \tilde{\alpha} \hat{\gamma} \rangle}\mathrm{Ad}^{-1}_{\tilde{h}}(M_{\bar w})
&=
\left[
P^T - \sigma\, \mathrm{Ad}_h^{-1}\mathrm{Ad}_{\tilde h}
\right]^{-1}
\left(
t s^{-1}  J_{\bar w} - P^T\tilde J_{\bar w}
\right),
\qquad
\mathcal{N}_{\bar w} = 0.
\end{align}
Finally, using \eqref{lax1} and \eqref{lax2}, we obtain the final form of the two-dimensional action, given by
\begin{align}
S_{2d}
&=
r_{+}
\int_{\Sigma}\mathrm{vol}_2 \,
\mathrm{Tr}\!\Bigg(
J_w \,
\frac{\langle \alpha \beta \rangle}{\langle \alpha \hat{\gamma} \rangle}
\,\mathrm{Ad}_h^{-1}(M_{\bar w})
-
 J_{\bar w}\,
\frac{\langle \alpha \beta \rangle}{\langle \alpha \gamma \rangle}
\,\mathrm{Ad}_{h}^{-1}(M_w)
\Bigg)
\nonumber \\[8pt]
&\quad
+
r_{-}
\int_{\Sigma}\mathrm{vol}_2 \,
\mathrm{Tr}\!\Bigg(
\tilde J_w \,
\frac{\langle \tilde{\alpha} \beta \rangle}{\langle \tilde{\alpha} \hat{\gamma} \rangle}
\,\mathrm{Ad}_{\tilde h}^{-1}(M_{\bar w})
-
\tilde J_{\bar w}\,
\frac{\langle \tilde{\alpha} \beta \rangle}{\langle \tilde{\alpha} \gamma \rangle}
\,\mathrm{Ad}_{\tilde h}^{-1}(M_w)
\Bigg)+ \textrm{WZ terms}\nonumber\\
=&
	\int_{\Sigma} \mathrm{vol}_2 \,
	\mathrm{Tr}\Big(
	r_{+} J_{w}\big(P^TU_{+}^{T}-U_{-}P\big)J_{\bar{w}}
	- r_{-} \tilde{J}_{w}\big(U_{+}^{T}P^T-PU_{-}\big)\tilde{J}_{\bar{w}}\nonumber
		\\&- 2t\,\sqrt{-r_{+} r_{-}}\,
	\tilde{J}_{w} U_{+}^{T} J_{\bar{w}}
	+ 2t^{-1}\,\sqrt{-r_{+} r_{-}}\,
	J_{w} U_{-} \tilde{J}_{\bar{w}}
	\Big)+\textrm{WZ terms},
\end{align}
which is exactly same as the two-dimensional field theory derived in \eqref{ift22}.

We have described how this 2d field theory with two fields includes the Yang--Baxter sigma model as a special case in \eqref{ybreduc}. We would like to address the question of when this 2d two-field theory is integrable in general. This requires additional investigation despite the fact that we have derived a flat Lax connection as part of the derivation from 4d CS, because such field theories with two fields may require more than one Lax connection to realise all nonlocal integrals of motion. Indeed, two-field models such as the one we derive here are expected to arise from coupled sigma models in a certain limit, as explained in \cite{Bassi:2019aaf}. In this limit,  two possible Lax operators can arise depending on how one takes the limit of the Lax operator of the original theory. Each of the resulting Lax operators loses complementary information about the parameters of the original theory (corresponding to poles of the twist function) in the limit. Hence, both Lax operators are expected to be necessary for integrability of the resulting theory. We expect that this is also the case for the two-field model we have derived. 

As explained in Section \eqref{semilcs}, when we specialise $\mathcal{O}$ to $R$ and $\sigma$ to $\frac{1+\eta c}{1-\eta c}$, the $\textrm{IFT}_4$ admits a semi-local symmetry, which in turn is expected to give rise to an additional 4d Lax operator called a $C$-Lax. Both the $B$-Lax and $C$-Lax can be symmetry reduced to two-dimensional Lax operators, as explained in \cite{Cole:2023umd}.
In general, we expect that whenever the $\textrm{IFT}_4$ derived in \eqref{ift4actionn} admits a semi-local symmetry, there ought to be an additional $C$-Lax, that further symmetry reduces to a second Lax pair in 2d, ensuring the integrability of the 2d model. It would be interesting to find explicit examples of such  semi-local symmetries that do not correspond to the $\textrm{IFT}_4^{\textrm{YB}}$ case of $\mathcal{O}=R$ and $\sigma=\frac{1+\eta c}{1-\eta c}$ that we considered in this work.

\section{Homogeneous Yang--Baxter model from the Chern--Simons route}
\label{sec:homogeneous_CS_route}

In this section we derive the homogeneous Yang--Baxter sigma model along the
Chern--Simons side of the diamond,
\begin{equation}
	\mathrm{hCS}_6
	\longrightarrow
	\mathrm{CS}_4
	\longrightarrow
	\mathrm{IFT}_2 .
\end{equation}
We start from the Yang--Baxter boundary condition obtained in
Section~\ref{6dCSto4dCS} after reducing the six-dimensional twistor theory to
four-dimensional Chern--Simons theory.  The homogeneous limit is obtained by
bringing the pair of simple-pole locations of the inhomogeneous Yang--Baxter
construction to a common point on the auxiliary $\mathbb{CP}^1$ fibre.  In
this limit the boundary condition is no longer expressed in terms of the
values of the Chern--Simons gauge field at two separated simple poles.
Instead, it relates the value of the gauge field at the coalesced double pole
to its first $\zeta$-coefficient there.

We work in the affine patch
\begin{equation}
	\zeta=\frac{\pi_{2'}}{\pi_{1'}}
\end{equation}
and choose
\begin{equation}
	\alpha_{A'}=(1,-a),
	\qquad
	\tilde\alpha_{A'}=(1,a),
	\qquad
	\beta_{A'}=(0,1),
	\qquad
	a=c\eta .
	\label{eq:hom_CS_affine_data}
\end{equation}
Thus $\pi=\alpha$ and $\pi=\tilde\alpha$ are located at $\zeta=-a$ and
$\zeta=a$, respectively.  The condition at the double pole $\pi=\beta$ is kept
fixed, namely
\begin{equation}
	\mathcal A_A\big|_{\pi=\beta}=0 .
\end{equation}
We also keep the zeroes of the meromorphic one-form $\omega$ fixed, as in the
inhomogeneous Yang--Baxter construction before taking the coalescence limit,
\begin{equation}
	\gamma:\zeta=1,
	\qquad
	\hat\gamma:\zeta=-1,
	\label{eq:hom_CS_zeroes}
\end{equation}
with representatives
\begin{equation}
	\gamma_{A'}=(1,1),
	\qquad
	\hat\gamma_{A'}=(1,-1).
\end{equation}
For these spinors one finds
\begin{equation}
	\frac{\langle\alpha\gamma\rangle}
	     {\langle\tilde\alpha\gamma\rangle}
	=
	\frac{1+a}{1-a},
	\qquad
	\frac{\langle\alpha\hat\gamma\rangle}
	     {\langle\tilde\alpha\hat\gamma\rangle}
	=
	\frac{1-a}{1+a}.
\end{equation}
Therefore the specialization \eqref{sigmachoice} gives
\begin{equation}
	\sigma
	=
	\frac{\langle\alpha\gamma\rangle}
	     {\langle\tilde\alpha\gamma\rangle}
	=
	\frac{\langle\tilde\alpha\hat\gamma\rangle}
	     {\langle\alpha\hat\gamma\rangle}
	=
	\frac{1+a}{1-a}.
	\label{eq:hom_CS_sigma}
\end{equation}
With this choice, the parameters defined in \eqref{seqn} and \eqref{teqn}
reduce to
\begin{equation}
	s=1,
	\qquad
	t=1 .
\end{equation}

We now reduce the six-dimensional boundary condition to the Chern--Simons
boundary condition.  The matching relation between the twistor connection and
the reduced Chern--Simons gauge field is
\begin{equation}
	A_w
	=
	-\frac{[\mathcal A\mu]}{\langle\pi\gamma\rangle},
	\qquad
	A_{\bar w}
	=
	-\frac{[\mathcal A\hat\mu]}{\langle\pi\hat\gamma\rangle}.
	\label{eq:hom_CS_matching}
\end{equation}
At $\pi=\alpha$ and $\pi=\tilde\alpha$, this gives
\begin{equation}
	[\mathcal A\mu]\big|_{\pi=\alpha}
	=
	-\langle\alpha\gamma\rangle A_w(-a),
	\qquad
	[\mathcal A\mu]\big|_{\pi=\tilde\alpha}
	=
	-\langle\tilde\alpha\gamma\rangle A_w(a).
\end{equation}
Substituting these expressions into the first equation in
\eqref{boundary conditions}, and using \eqref{eq:hom_CS_sigma}, the spinor
factors cancel.  Hence
\begin{equation}
	(R-c)A_w(-a)
	=
	(R+c)A_w(a).
\end{equation}
The hatted component gives the same relation for $A_{\bar w}$ because
\begin{equation}
	\sigma^{-1}
	=
	\frac{\langle\alpha\hat\gamma\rangle}
	     {\langle\tilde\alpha\hat\gamma\rangle}.
\end{equation}
Therefore both components obey
\begin{equation}
	(R-c)A_i(-a)
	=
	(R+c)A_i(a),
	\qquad
	i=w,\bar w .
	\label{eq:hom_CS_finite_boundary_relation}
\end{equation}
This is the Yang--Baxter boundary condition \eqref{4dBC}, written in the
present affine parametrisation with $\alpha_+=-a$ and $\alpha_-=a$.

We now take the homogeneous scaling limit.  The combination kept fixed is
\begin{equation}
	R_{\rm h}
	=
	\eta R
	=
	\frac{a}{c}R ,
	\label{eq:hom_CS_Rh_def}
\end{equation}
while $a\to0$.  Equivalently,
\begin{equation}
	R=\frac{c}{a}R_{\rm h}.
\end{equation}
Thus
\begin{equation}
	R-c=\frac{c}{a}(R_{\rm h}-a),
	\qquad
	R+c=\frac{c}{a}(R_{\rm h}+a).
\end{equation}
The boundary relation before the homogeneous limit,
\eqref{eq:hom_CS_finite_boundary_relation}, becomes
\begin{equation}
	(R_{\rm h}-a)A_i(-a)
	=
	(R_{\rm h}+a)A_i(a),
	\qquad
	i=w,\bar w .
	\label{eq:hom_CS_scaled_BC}
\end{equation}
Expanding around the coalesced point $\zeta=0$,
\begin{align}
	A_i(-a)
	&=
	A_i(0)
	-
	a\,\partial_\zeta A_i(0)
	+
	O(a^2),
	\\
	A_i(a)
	&=
	A_i(0)
	+
	a\,\partial_\zeta A_i(0)
	+
	O(a^2),
\end{align}
and substituting into \eqref{eq:hom_CS_scaled_BC}, the leading terms cancel.
The first non-trivial order gives
\begin{equation}
	A_i(0)
	+
	R_{\rm h}\partial_\zeta A_i(0)
	=
	0,
	\qquad
	i=w,\bar w .
	\label{eq:hom_CS_double_pole_BC}
\end{equation}
This is the homogeneous boundary condition at the double pole.  Here
$\partial_\zeta A_i(0)$ is the first $\zeta$-coefficient in the expansion of
the Chern--Simons gauge field around the coalesced pole.

The same relation follows directly in six-dimensional variables.  Let
\begin{equation}
	B(\zeta)=[\mathcal A\mu](\zeta).
\end{equation}
The homogeneous limit of the first equation in \eqref{boundary conditions}
gives
\begin{equation}
	B(0)
	+
	R_{\rm h}\partial_\zeta B(0)
	+
	R_{\rm h}B(0)
	=
	0.
	\label{eq:hom_CS_B_condition}
\end{equation}
Using
\begin{equation}
	B(\zeta)
	=
	-\langle\pi\gamma\rangle A_w(\zeta),
	\qquad
	\langle\pi\gamma\rangle=\zeta-1,
\end{equation}
we have
\begin{equation}
	B(\zeta)=(1-\zeta)A_w(\zeta).
\end{equation}
Therefore
\begin{equation}
	B(0)=A_w(0),
	\qquad
	\partial_\zeta B(0)
	=
	-A_w(0)+\partial_\zeta A_w(0).
\end{equation}
Substituting these expressions into \eqref{eq:hom_CS_B_condition} gives
\begin{equation}
	A_w(0)
	+
	R_{\rm h}\partial_\zeta A_w(0)
	=
	0.
\end{equation}
The hatted component works in the same way.  Since
\begin{equation}
	[\mathcal A\hat\mu](\zeta)
	=
	-\langle\pi\hat\gamma\rangle A_{\bar w}(\zeta),
	\qquad
	\langle\pi\hat\gamma\rangle=\zeta+1,
\end{equation}
the homogeneous limit of the second equation in \eqref{boundary conditions}
gives
\begin{equation}
	A_{\bar w}(0)
	+
	R_{\rm h}\partial_\zeta A_{\bar w}(0)
	=
	0.
\end{equation}
Thus the twistor boundary condition and the reduced Chern--Simons boundary
condition give the same double-pole relation.

The meromorphic one-form of the reduced four-dimensional Chern--Simons theory
is inherited from the contraction formula \eqref{redlag}.  With the
normalisation used there,
\begin{equation}
	\omega_a
	=
	-
	K
	\frac{
	\langle\pi\gamma\rangle
	\langle\pi\hat\gamma\rangle
	}{
	\langle\pi\alpha\rangle
	\langle\pi\tilde\alpha\rangle
	\langle\pi\beta\rangle^2
	}
	d\zeta .
	\label{eq:hom_CS_omega_a_general}
\end{equation}
In the affine patch,
\begin{equation}
	\langle\pi\beta\rangle=1,
	\qquad
	\langle\pi\gamma\rangle=\zeta-1,
	\qquad
	\langle\pi\hat\gamma\rangle=\zeta+1,
\end{equation}
and
\begin{equation}
	\langle\pi\alpha\rangle=\zeta+a,
	\qquad
	\langle\pi\tilde\alpha\rangle=\zeta-a.
\end{equation}
Therefore
\begin{equation}
	\omega_a
	=
	-
	K
	\frac{(\zeta-1)(\zeta+1)}
	     {(\zeta+a)(\zeta-a)}
	d\zeta
	=
	K
	\frac{1-\zeta^2}{\zeta^2-a^2}
	d\zeta .
	\label{eq:hom_CS_omega_a}
\end{equation}
Taking $a\to0$ gives
\begin{equation}
	\omega_{\rm h}
	=
	K
	\left(
	\frac{1}{\zeta^2}-1
	\right)d\zeta .
	\label{eq:hom_CS_omega_h}
\end{equation}
Thus the simple poles at $\zeta=\pm a$ combine into a double pole at
$\zeta=0$, while the zeroes at $\zeta=\pm1$ remain fixed.

The algebraic Yang--Baxter condition degenerates in parallel.  Substituting
$R=(c/a)R_{\rm h}$ into the modified classical Yang--Baxter equation gives
\begin{equation}
	[R_{\rm h}x,R_{\rm h}y]
	-
	R_{\rm h}
	\left(
	[R_{\rm h}x,y]+[x,R_{\rm h}y]
	\right)
	=
	-a^2[x,y].
\end{equation}
Taking $a\to0$, one obtains
\begin{equation}
	[R_{\rm h}x,R_{\rm h}y]
	-
	R_{\rm h}
	\left(
	[R_{\rm h}x,y]+[x,R_{\rm h}y]
	\right)
	=
	0.
	\label{eq:hom_CS_CYBE}
\end{equation}
Thus $R_{\rm h}$ satisfies the homogeneous classical Yang--Baxter equation.
Equivalently, the double-pole boundary condition
\eqref{eq:hom_CS_double_pole_BC} selects the graph of $-R_{\rm h}$ in the
space of boundary data $(A_i(0),\partial_\zeta A_i(0))$.  Closure of this
boundary algebra is equivalent to \eqref{eq:hom_CS_CYBE}, while
$R_{\rm h}^t=-R_{\rm h}$ gives isotropy of the corresponding boundary
pairing.

We now localise the four-dimensional Chern--Simons theory to a
two-dimensional integrable field theory.  The derivation is well-known from
\cite{Delduc:2019whp}, and we shall only quote the result:
\begin{equation}
	S_{\mathrm{IFT}_2}^{\rm hom}
	=
	2K
	\int_\Sigma\mathrm{vol}_2\,
	\Tr\left[
	J_w(1-R_{\rm h})^{-1}J_{\bar w}
	\right].
	\label{eq:hom_CS_IFT2_wbarw}
\end{equation}
Using
\begin{equation}
	J_+=J_w,
	\qquad
	J_-=J_{\bar w},
\end{equation}
this becomes
\begin{equation}
	S_{\mathrm{IFT}_2}^{\rm hom}
	=
	2K
	\int_\Sigma\mathrm{vol}_2\,
	\Tr\left[
	J_+(1-R_{\rm h})^{-1}J_-
	\right].
	\label{eq:hom_CS_IFT2_final}
\end{equation}
Equivalently, after defining $\kappa=2K$, we obtain
\begin{equation}
	S_{\rm YB}^{\rm hom}
	=
	\kappa
	\int_\Sigma\mathrm{vol}_2\,
	\Tr\left[
	J_+(1-R_{\rm h})^{-1}J_-
	\right].
	\label{eq:hom_CS_YB_final}
\end{equation}

The same result is obtained as a consistency check by taking the homogeneous
limit of the inhomogeneous Yang--Baxter action derived in
Section~\ref{ybreduc}. The inhomogeneous Yang--Baxter action is
\begin{equation}
	S_{\mathrm{IFT}_2}
	=
	C_1
	\int_\Sigma\mathrm{vol}_2\,
	\Tr\left[
	J_+
	\frac{1}{1-\eta R}
	J_-
	\right],
	\qquad
	C_1=4r_+c\eta .
	\label{eq:hom_CS_inhom_action}
\end{equation}
In the homogeneous scaling \eqref{eq:hom_CS_Rh_def},
\begin{equation}
	1-\eta R=1-R_{\rm h}.
\end{equation}
Furthermore,
\begin{equation}
	r_+
	=
	K
	\frac{
	\langle\alpha\gamma\rangle
	\langle\alpha\hat\gamma\rangle
	}{
	\langle\alpha\tilde\alpha\rangle
	\langle\alpha\beta\rangle^2
	}.
\end{equation}
For the affine representatives used above,
\begin{equation}
	\langle\alpha\gamma\rangle=-(1+c\eta),
	\qquad
	\langle\alpha\hat\gamma\rangle=1-c\eta,
	\qquad
	\langle\alpha\tilde\alpha\rangle=-2c\eta,
	\qquad
	\langle\alpha\beta\rangle=1.
\end{equation}
Therefore
\begin{equation}
	r_+
	=
	K\frac{1-c^2\eta^2}{2c\eta}.
\end{equation}
It follows that
\begin{equation}
	C_1
	=
	4r_+c\eta
	=
	2K(1-c^2\eta^2)
	\longrightarrow
	2K
	\qquad
	\mathrm{as}
	\qquad
	\eta\to0.
\end{equation}
Thus \eqref{eq:hom_CS_inhom_action} reduces precisely to
\eqref{eq:hom_CS_IFT2_final}.

The homogeneous Yang--Baxter model is therefore obtained from the same
twistor-space construction by the coalescence
\begin{equation}
	\alpha:\zeta=-c\eta,
	\qquad
	\tilde\alpha:\zeta=c\eta
	\qquad
	\longrightarrow
	\qquad
	\zeta=0 .
\end{equation}
The combination
\begin{equation}
	R_{\rm h}=\eta R
\end{equation}
survives in the first $\zeta$-coefficient of the reduced Chern--Simons gauge
field at the double pole.  At the same time, the modified classical
Yang--Baxter equation degenerates to the homogeneous classical Yang--Baxter
equation, and the localised two-dimensional theory becomes the homogeneous
Yang--Baxter sigma model.  In the next section we reproduce the same result
by following the other side of the diamond, namely by taking the homogeneous
limit of $\mathrm{IFT}_4$ and then reducing to two dimensions.

\section{Homogeneous limit and symmetry reduction}
\label{sec:homogeneous_limit_and_reduction}

In Section~\ref{sec:homogeneous_CS_route} we obtained the homogeneous
Yang--Baxter sigma model by first reducing the twistor theory to
four-dimensional Chern--Simons theory and then localising the resulting
Chern--Simons theory to two dimensions.  We now follow the other side of the
diamond.  We take the homogeneous limit directly in the four-dimensional
integrable field theory \eqref{ift4actionn}, and then reduce the resulting
homogeneous $\mathrm{IFT}_4$ to two dimensions.  The final answer agrees with
\eqref{eq:hom_CS_IFT2_final}.

\subsection{Homogeneous limit of the four-dimensional field theory}
\label{subsec:homogeneous_limit_IFT4}

We start from the four-dimensional action before the homogeneous limit,
\eqref{ift4actionn},
\begin{align}
	S_{\mathrm{IFT}_4}
	=
	&
	\frac{K}{\langle\alpha\tilde\alpha\rangle}
	\int_{\mathbb E^4}\mathrm{vol}_4\,
	\Tr\Big[
	U_+(Pj-\sigma\tilde j)
	(\hat j-\Lambda^T\hat{\tilde j})
	\nonumber\\
	&\hspace{7em}
	-
	U_-(P\hat j-\sigma^{-1}\hat{\tilde j})
	(j-\Lambda^T\tilde j)
	\Big]
	+
	S_{\mathrm{WZ}_4}.
	\label{eq:hom_IFT4_start}
\end{align}
Here $P$, $U_\pm$ and $\Lambda$ are defined in \eqref{Operators}.  In the
Yang--Baxter specialization,
\begin{equation}
	P=\frac{R-c}{R+c},
	\qquad
	U_\pm=(P-\sigma^{\pm1}\Lambda)^{-1},
	\qquad
	\Lambda=\operatorname{Ad}_{\tilde h}^{-1}\operatorname{Ad}_{h}.
	\label{eq:hom_IFT4_PU}
\end{equation}
The currents $j,\hat j,\tilde j,\hat{\tilde j}$ are those in
\eqref{currents}.

We now use the same coalescence data as in
Section~\ref{sec:homogeneous_CS_route}.  Thus the two simple-pole locations
are $\zeta=-a$ and $\zeta=a$, with $a=c\eta$, and the coalesced point is
denoted by $\alpha_0$.  In spinor notation this is
\begin{equation}
	\alpha_{A'}=\alpha_{0A'}-a\beta_{A'},
	\qquad
	\tilde\alpha_{A'}=\alpha_{0A'}+a\beta_{A'}.
	\label{eq:hom_IFT4_alpha_expand}
\end{equation}
After raising the primed index,
\begin{equation}
	\alpha^{A'}=\alpha_0^{A'}-a\beta^{A'},
	\qquad
	\tilde\alpha^{A'}=\alpha_0^{A'}+a\beta^{A'}.
	\label{eq:hom_IFT4_alpha_raised}
\end{equation}
The parameter $\sigma$ is given by \eqref{eq:hom_CS_sigma}, and therefore
\begin{equation}
	\sigma=1+2a+O(a^2),
	\qquad
	\sigma^{-1}=1-2a+O(a^2).
	\label{eq:hom_IFT4_sigma_expand}
\end{equation}
The homogeneous Yang--Baxter operator is the fixed combination
\eqref{eq:hom_CS_Rh_def},
\begin{equation}
	R_{\rm h}=\eta R=\frac{a}{c}R .
\end{equation}
Equivalently,
\begin{equation}
	R=\frac{c}{a}R_{\rm h},
	\qquad
	P=(R_{\rm h}-a)(R_{\rm h}+a)^{-1}.
	\label{eq:hom_IFT4_P_expand}
\end{equation}
The operator $R_{\rm h}$ is skew-symmetric and satisfies the homogeneous
classical Yang--Baxter equation \eqref{eq:hom_CS_CYBE}.

The standard homogeneous Yang--Baxter model arises from the diagonal
coalescence branch,
\begin{equation}
	h=\tilde h.
	\label{eq:hom_IFT4_diagonal_branch}
\end{equation}
On this branch,
\begin{equation}
	\Lambda=1,
	\qquad
	\Lambda^T=1,
\end{equation}
so that
\begin{equation}
	U_+=(P-\sigma)^{-1},
	\qquad
	U_-=(P-\sigma^{-1})^{-1}.
\end{equation}
This branch is the four-dimensional counterpart of the one-field
homogeneous Yang--Baxter model.  More general double-pole limits may keep
extra edge data, but they are not needed for the standard homogeneous model.

Using \eqref{eq:hom_IFT4_alpha_raised}, the currents have the following small-$a$ expansion
\begin{align}
	j
	&=
	j_0-a\,j_\beta,
	&
	\tilde j
	&=
	j_0+a\,j_\beta,
\nonumber
	\\
	\hat j
	&=
	\hat j_0-a\,\hat j_\beta,
	&
	\hat{\tilde j}
	&=
	\hat j_0+a\,\hat j_\beta,
	\label{eq:hom_IFT4_jhat_expand}
\end{align}
where
\begin{align}
	j_0
	&=
	\mu^A\alpha_0^{A'}h^{-1}\partial_{AA'}h,
	&
	j_\beta
	&=
	\mu^A\beta^{A'}h^{-1}\partial_{AA'}h,\nonumber\\
	\hat j_0
	&=
	\hat\mu^A\alpha_0^{A'}h^{-1}\partial_{AA'}h,
	&
	\hat j_\beta
	&=
	\hat\mu^A\beta^{A'}h^{-1}\partial_{AA'}h.
	\label{eq:hom_IFT4_jhat0_jhatbeta}
\end{align}
Therefore
\begin{equation}
	j-\tilde j=-2a\,j_\beta,
	\qquad
	\hat j-\hat{\tilde j}=-2a\,\hat j_\beta.
	\label{eq:hom_IFT4_current_difference}
\end{equation}
At the same time,
\begin{equation}
	\langle\alpha\tilde\alpha\rangle=2a.
	\label{eq:hom_IFT4_denominator}
\end{equation}
Thus the factor $1/\langle\alpha\tilde\alpha\rangle$ in
\eqref{eq:hom_IFT4_start} is compensated by the difference of the two
currents. This cancellation gives a well-defined homogeneous
four-dimensional action.

We now evaluate the two combinations appearing in \eqref{eq:hom_IFT4_start}.
First define
\begin{equation}
	B_+(a)=U_+(Pj-\sigma\tilde j).
\end{equation}
Using \eqref{eq:hom_IFT4_P_expand}, this equation is equivalent to
\begin{equation}
	\left[
	(R_{\rm h}-a)-\sigma(R_{\rm h}+a)
	\right]B_+(a)
	=
	(R_{\rm h}-a)j
	-
	\sigma(R_{\rm h}+a)\tilde j.
	\label{eq:hom_IFT4_Bplus_eq}
\end{equation}
Substituting \eqref{eq:hom_IFT4_sigma_expand} and
\eqref{eq:hom_IFT4_jhat_expand}, and keeping the first non-trivial order in
$a$, gives
\begin{equation}
	B_+(a)
	\longrightarrow
	b_{\rm h}
	=
	j_0
	+
	R_{\rm h}(1+R_{\rm h})^{-1}j_\beta.
	\label{eq:hom_IFT4_bh}
\end{equation}

Similarly, define
\begin{equation}
	B_-(a)=U_-(P\hat j-\sigma^{-1}\hat{\tilde j}).
\end{equation}
Using \eqref{eq:hom_IFT4_P_expand}, this becomes
\begin{equation}
	\left[
	(R_{\rm h}-a)-\sigma^{-1}(R_{\rm h}+a)
	\right]B_-(a)
	=
	(R_{\rm h}-a)\hat j
	-
	\sigma^{-1}(R_{\rm h}+a)\hat{\tilde j}.
	\label{eq:hom_IFT4_Bminus_eq}
\end{equation}
Substituting \eqref{eq:hom_IFT4_sigma_expand} and
\eqref{eq:hom_IFT4_jhat_expand} gives
\begin{equation}
	B_-(a)
	\longrightarrow
	\hat b_{\rm h}
	=
	\hat j_0
	+
	R_{\rm h}(1-R_{\rm h})^{-1}\hat j_\beta.
	\label{eq:hom_IFT4_bhhat}
\end{equation}

Using \eqref{eq:hom_IFT4_current_difference},
\eqref{eq:hom_IFT4_denominator}, \eqref{eq:hom_IFT4_bh} and
\eqref{eq:hom_IFT4_bhhat}, the kinetic part of the action becomes
\begin{equation}
	S_{\rm kin}^{\rm hom}
	=
	K
	\int_{\mathbb E^4}\mathrm{vol}_4\,
	\Tr\left[
	\hat b_{\rm h}\,j_\beta
	-
	b_{\rm h}\,\hat j_\beta
	\right].
	\label{eq:hom_IFT4_kinetic}
\end{equation}

The Wess--Zumino term also gives a well-defined homogeneous limit. On the diagonal
branch,
\begin{equation}
	[j,\hat j]-[\tilde j,\hat{\tilde j}]
	=
	-2a
	\left(
	[j_\beta,\hat j_0]
	+
	[j_0,\hat j_\beta]
	\right).
\end{equation}
Therefore
\begin{align}
	S_{\rm WZ}^{\rm hom}
	=
	-
	K
	\int_{\mathbb E^4\times[0,1]}
	\mathrm{vol}_4\wedge d\rho\,
	\Tr\left[
	h^{-1}\partial_\rho h
	\left(
	[j_\beta,\hat j_0]
	+
	[j_0,\hat j_\beta]
	\right)
	\right].
	\label{eq:hom_IFT4_WZ}
\end{align}
The homogeneous four-dimensional theory obtained from the $\mathrm{IFT}_4$
side of the diamond is therefore
\begin{equation}
	S_{\mathrm{IFT}_4}^{\rm hom}
	=
	S_{\rm kin}^{\rm hom}
	+
	S_{\rm WZ}^{\rm hom}.
	\label{eq:hom_IFT4_action}
\end{equation}

\subsection{Reduction to two dimensions}
\label{subsec:hom_IFT4_to_IFT2}

We now reduce \eqref{eq:hom_IFT4_action} to two dimensions.  The reduction is
the same as in Section~\ref{ybreduc}: the field is independent of the two
directions removed by the reduction,
\begin{equation}
	\partial_z h=0,
	\qquad
	\partial_{\bar z}h=0.
\end{equation}
The remaining coordinates are
\begin{equation}
	\sigma^+=w,
	\qquad
	\sigma^-=\bar w,
\end{equation}
and we define
\begin{equation}
	J_+=h^{-1}\partial_+h,
	\qquad
	J_-=h^{-1}\partial_-h,
	\qquad
	\partial_+=\partial_w,
	\qquad
	\partial_-=\partial_{\bar w}.
\end{equation}

For any fixed primed spinor $X^{A'}$, we use the spinor decomposition
\eqref{symmetry reduction spinor}.  Together with the reduction constraints,
it gives
\begin{equation}
	\mu^A X^{A'}h^{-1}\partial_{AA'}h
	\longrightarrow
	\langle X\gamma\rangle J_+,
	\qquad
	\hat\mu^A X^{A'}h^{-1}\partial_{AA'}h
	\longrightarrow
	\langle X\hat\gamma\rangle J_-.
	\label{eq:hom_IFT4_reduction_rules}
\end{equation}
It is convenient to introduce
\begin{equation}
	A=\langle\alpha_0\gamma\rangle,
	\qquad
	B=\langle\beta\gamma\rangle,
	\qquad
	C=\langle\alpha_0\hat\gamma\rangle,
	\qquad
	D=\langle\beta\hat\gamma\rangle .
	\label{eq:hom_IFT4_ABCD}
\end{equation}
Then
\begin{align}
j_0\longrightarrow A\,J_+, \qquad j_\beta\longrightarrow B\,J_+, \qquad \hat j_0\longrightarrow C\,J_-, \qquad\hat j_\beta\longrightarrow D\,J_-.
\end{align}
Thus
\begin{align}
	b_{\rm h}
	\longrightarrow
	\left[
	A
	+
	B\,R_{\rm h}(1+R_{\rm h})^{-1}
	\right]J_+,\qquad
	\hat b_{\rm h}
	\longrightarrow
	\left[
	C
	+
	D\,R_{\rm h}(1-R_{\rm h})^{-1}
	\right]J_-.
\label{eq:hom_IFT4_bhat_reduced}
\end{align}

The kinetic term \eqref{eq:hom_IFT4_kinetic} reduces to
\begin{equation}
	S_{\rm kin}^{\rm hom}
	\longrightarrow
	K
	\int_{\Sigma}\mathrm{vol}_2\,
	\Tr\left[
	J_+\,\mathcal M_{\rm h}\,J_-
	\right],
	\label{eq:hom_IFT4_kin_reduced}
\end{equation}
where the operator $\mathcal M_{\rm h}$ is obtained by using
$R_{\rm h}^t=-R_{\rm h}$:
\begin{equation}
	\mathcal M_{\rm h}
	=
	(BC-AD)
	+
	2BD\,R_{\rm h}(1-R_{\rm h})^{-1}.
	\label{eq:hom_IFT4_Mh}
\end{equation}

The Wess--Zumino term reduces as
\begin{equation}
	[j_\beta,\hat j_0]+[j_0,\hat j_\beta]
	\longrightarrow
	(BC+AD)[J_+,J_-].
\end{equation}
Hence
\begin{align}
	S_{\rm WZ}^{\rm hom}
	\longrightarrow
	-
	K(BC+AD)
	\int_{\Sigma\times[0,1]}
	\mathrm{vol}_2\wedge d\rho\,
	\Tr\left[
	h^{-1}\partial_\rho h\,[J_+,J_-]
	\right].
\end{align}
Equivalently, defining
\begin{equation}
	\mathcal L_{\rm WZ}(h)
	=
	-
	\int_0^1d\rho\,
	\Tr\left[
	h^{-1}\partial_\rho h\,[J_+,J_-]
	\right],
\end{equation}
the reduced homogeneous action is
\begin{align}
	S_{\mathrm{IFT}_2}^{\rm hom}
	=
	K
	\int_{\Sigma}\mathrm{vol}_2\,
	\Tr\left[
	J_+\,\mathcal M_{\rm h}\,J_-
	\right]
	+
	K(BC+AD)
	\int_{\Sigma}\mathrm{vol}_2\,\mathcal L_{\rm WZ}(h).
	\label{eq:hom_IFT4_general_reduced_action}
\end{align}

We finally impose the same zeroes as in \eqref{eq:hom_CS_zeroes}.  With the
affine convention of Section~\ref{sec:homogeneous_CS_route}, the required
spinor products are
\begin{equation}
	BC=1,
	\qquad
	AD=-1,
	\qquad
	BD=1,
	\qquad
	BC+AD=0.
	\label{eq:hom_IFT4_standard_products}
\end{equation}
The Wess--Zumino term therefore vanishes.  The kinetic operator becomes
\begin{align}
	\mathcal M_{\rm h}
	&=
	(BC-AD)
	+
	2BD\,R_{\rm h}(1-R_{\rm h})^{-1}
	\nonumber\\
	&=
	2
	+
	2R_{\rm h}(1-R_{\rm h})^{-1}
	\nonumber\\
	&=
	2(1-R_{\rm h})^{-1}.
\end{align}
Substituting this into \eqref{eq:hom_IFT4_general_reduced_action}, we obtain
\begin{equation}
	S_{\mathrm{IFT}_2}^{\rm hom}
	=
	2K
	\int_{\Sigma}\mathrm{vol}_2\,
	\Tr\left[
	J_+(1-R_{\rm h})^{-1}J_-
	\right].
	\label{eq:hom_IFT4_final_reduced_action}
\end{equation}
This is exactly the result obtained from the Chern--Simons route in
\eqref{eq:hom_CS_IFT2_final}.  After the coupling redefinition $\kappa=2K,$
we get
\begin{equation}
	S_{\rm YB}^{\rm hom}
	=
	\kappa
	\int_{\Sigma}\mathrm{vol}_2\,
	\Tr\left[
	J_+(1-R_{\rm h})^{-1}J_-
	\right].
	\label{eq:hom_IFT4_YB_final}
\end{equation}

Thus the homogeneous limit is compatible with both routes of the diamond.  On
the Chern--Simons side it appears as the double-pole boundary condition
\eqref{eq:hom_CS_double_pole_BC}. On the four-dimensional field-theory side it appears as the one-field
homogeneous action \eqref{eq:hom_IFT4_action}. After
symmetry reduction, both routes give the same homogeneous Yang--Baxter sigma
model.

\section{Conclusion}
We have derived a new two-dimensional field theory with action \eqref{ift22} from 4d Chern--Simons theory and its associated twistor space Chern--Simons theory, in particular, by starting with the action \eqref{6daction} and boundary conditions \eqref{boundary conditions}. Its origin from 4d Chern--Simons theory suggests that is integrable, given that the 4d equations of motion include the flatness of a gauge field that can be interpreted as a Lax connection. It would be interesting to derive the flatness of the Lax connection predicted by 4d 
Chern--Simons theory explicitly using the equations of motion of the two-dimensional theory.

The two-dimensional field theory was shown to be related via symmetry reduction to a novel 4d field theory depending on a skew-symmetric operator, $\mathcal{O}$.
We showed that by choosing the operator $\mathcal{O}$ to be a solution $R$ of the modified classical Yang--Baxter equation, the 4d theory gains a semi-local symmetry. Symmetry reducing this 4d Yang--Baxter sigma model results in the familiar 2d Yang--Baxter sigma model associated with the modified classical Yang--Baxter sigma model.  Finally, we showed how the homogenous Yang--Baxter sigma model associated with ordinary classical Yang--Baxter sigma model arises in the limit where simple poles in the meromorphic 3-form of the twistor space Chern--Simons theory coalesce to become a double pole.

There are several future directions that may be pursued based on this work. Firstly, a natural generalisation of our results would be the derivation of the Yang--Baxter sigma model with Wess--Zumino term from twistor space. Secondly, given that the twistor space origins of both the Yang--Baxter sigma model and $\lambda$-model \cite{Cole:2023umd} are known, it would be interesting to investigate the twistor space origin of the Poisson--Lie T-duality between 
these theories.  Going beyond classical aspects, it may also be interesting to derive the quantisation of both the 4d and 2d Yang--Baxter sigma model from twistor space Chern--Simons theory, leveraging known results on quantising 6d holomorphic field theories by Costello \cite{Costello:2021bah}.

%summary - new 2d ift, might be integrable . showed choosing O to be R led to semilocal symmetry
%1. Future ideas - poisson-lie t-duality, quantization in 6d, yb sigma model with wess-zumino term. tst duality origin in 6d?
\appendix

\section{Twistor Conventions } \label{ap A}

The twistor space, $\mathbb{P T}$, of complexified Minkowski space, $\mathbb{C M}^4$, is the total space of the holomorphic vector bundle
\ie \label{twistorbundle}
\mathcal{O}(1) \oplus \mathcal{O}(1) \rightarrow \mathbb{C P}^1,
\fe
which can be endowed with homogeneous coordinates $Z^\alpha=\left(\omega^A, \pi_{A^{\prime}}\right)$ defined with respect to the equivalence relation $Z^\alpha \sim t Z^\alpha$ for $t \in \mathbb{C}^*$. Here, the unprimed (primed) indices $A,B,C,\ldots$ ($A',B',C',\ldots$) label elements of the right-handed (left-handed) spin bundle on $\mathbb{C M}^4$.
In other words, we have $\omega^A=\left(\omega^0, \omega^1\right)$ and $\pi_{A^{\prime}}=\left(\pi_{0^{\prime}}, \pi_{1^{\prime}}\right)$, that are, respectively, coordinates on the base and fibre of \eqref{twistorbundle}.  The points $x^{A A^{\prime}} \in \mathbb{C M}^4$ are in bijection with holomorphic lines, that is, 
\ie 
\iota_x: \mathbb{C P}_x^1 \hookrightarrow \mathbb{P T}, \quad \pi_{A^{\prime}} \mapsto\left(\omega^A, \pi_{A^{\prime}}\right)=\left(x^{A B^{\prime}} \pi_{B^{\prime}}, \pi_{A^{\prime}}\right).
\fe

We shall impose the reality structure associated with 4d Euclidean space $\mathbb{E}^4$. In Euclidean signature, primed (unprimed) spinors are mapped to primed (unprimed) spinors, that is, for $\omega^A=(\omega^0,\omega^1)$ and $\pi_{A'}=(\pi_0,\pi_1)$, we have
\ie \begin{aligned}\\
&\omega^A \rightarrow \hat{\omega}^A=\left(-\overline{\omega^1}, \overline{\omega^0}\right) \quad \text { and } \quad \pi_{A^{\prime}} \rightarrow \hat{\pi}_{A^{\prime}}=\left(-\overline{\pi_{1^{\prime}}}, \overline{\pi_{0^{\prime}}}\right).
\end{aligned}\fe
Having imposed the Euclidean reality structure, primed indices label elements of the left-handed spin bundle over $\mathbb{E}^4$, while unprimed indices  label elements of the right-handed spin bundle over $\mathbb{E}^4$.

Inner products for left-handed and right-handed spinors shall be defined respectively as 
\begin{equation}
\begin{aligned}
& \langle\pi \hat{\pi}\rangle=\pi^{A^{\prime}} \hat{\pi}_{A^{\prime}}=\varepsilon_{A^{\prime} B^{\prime}} \pi^{A^{\prime}} \hat{\pi}^{B^{\prime}}=\varepsilon^{A^{\prime} B^{\prime}} \pi_{B^{\prime}} \hat{\pi}_{A^{\prime}},\\
& {[\omega \hat{\omega}]=\omega^A \hat{\omega}_A=\varepsilon_{A B} \omega^A \hat{\omega}^B=\varepsilon^{A B} \omega_B \hat{\omega}_A} .
\end{aligned}
\end{equation}

We utilise a frame of holomorphic $(0,1)$-forms adapted to the non-holomorphic coordinates $\left(x^{A A^{\prime}}, \pi_{A^{\prime}}\right)$ on $\mathbb{P} \mathbb{T}$ :
\ie 
\bar{e}^0=\frac{\langle\mathrm{d} \hat{\pi} \hat{\pi}\rangle}{\|\pi\|^4} \in \Omega^{0,1}(\mathbb{P T}, \mathcal{O}(-2)), \quad \hat{e}^A=\frac{\mathrm{d} x^{A A^{\prime}} \hat{\pi}_{A^{\prime}}}{\|\pi\|^2} \in \Omega^{0,1}(\mathbb{P T}, \mathcal{O}(-1)),
\fe
 where 
$\|\pi\|^2=\pi_{A^{\prime}} \hat{\pi}^{A^{\prime}}$.
The dual frame of $(0,1)$-vectors is given by
\ie 
\bar{\partial}_0=\|\pi\|^2 \pi^{A^{\prime}} \frac{\partial}{ \partial \hat{\pi}^{A^{\prime}}} , \quad \hat{\partial}_A=\pi^{A^{\prime}} \frac{\partial}{ \partial x^{A A^{\prime}}} =\pi^{A^{\prime}} \partial_{{A A^{\prime}}}.
\fe
We shall also use 
\ie 
e^0=\langle\mathrm{d} \pi \pi\rangle.
\fe 

%It is often useful to employ an inhomogeneous coordinate, denoted $\zeta$, on the twistor sphere, which is related to $\pi^A$ using some right-handed spinors $\alpha$ and $\beta$ as 
%\ie \label{inhommg}
%\pi^A\sim\alpha^A- z \beta^A
%\fe 
%(the symbol $\sim$ here denotes equivalence up to an overall scale factor, which is equal to $\langle \pi \beta \rangle $),
%where 
%$
%\zeta=\frac{\langle \pi \alpha\rangle }{\langle \pi \beta \rangle }$, with
%\ie 
%\mathrm{d} \zeta=\frac{e^0}{\langle\pi \beta\rangle^2}, \quad \frac{\partial}{\partial \zeta}=\langle\pi \beta\rangle^2 \partial_0.
%\fe 
%In addition,
%\ie 
%\bar{e}^0=\frac{1}{\langle \pi \beta\rangle ^2}\frac{\textrm{d} \bar{\zeta}}{(1+|\zeta|^2)^2}, \quad  \bar{\partial}_0=\langle \pi \beta\rangle ^2 (1+|\zeta|^2)^2 \frac{\partial}{\partial \bar{\zeta}}.
%\fe

\section{Some Useful Identities} \label{ap B}
We would like to first derive the identity
\begin{align}\label{adjoint commutator}
\mathrm{Ad}_{h^{-1}}\!\left[ B_{A}, B^{A} \right]
=
\left[
\mathrm{Ad}_{h^{-1}} B_{A},
\mathrm{Ad}_{h^{-1}} B^{A}
\right].
\end{align}
This is simply the statement that the adjoint action defines a Lie-algebra automorphism, i.e.
\begin{align}\label{eq:adjoint-step}
\mathrm{Ad}_{h}\!\left([X,Y]\right)
=
\left[\mathrm{Ad}_{h} X, \mathrm{Ad}_{h} Y\right].
\end{align}
We begin by recalling the definition of the commutator,
\begin{equation}
[X,Y] := XY - YX.
\end{equation}
Then
\begin{align}
\mathrm{Ad}_{h}[X,Y]
&= h[X,Y]h^{-1} \nonumber \\
&= h(XY - YX)h^{-1} \nonumber \\
&= hXYh^{-1} - hYXh^{-1}.
\end{align}
We now insert the identity $1 = hh^{-1}$ at appropriate places in order to factor each term
\begin{align}
hXYh^{-1}
&= (hXh^{-1})(hYh^{-1}), \\
hYXh^{-1}
&= (hYh^{-1})(hXh^{-1}).
\end{align}
Therefore, \eqref{eq:adjoint-step} follows from
\begin{align}
\mathrm{Ad}_{h}[X,Y]
&= (hXh^{-1})(hYh^{-1}) - (hYh^{-1})(hXh^{-1}) \notag \\
&= \big[\, hXh^{-1},\, hYh^{-1} \,\big] \label{eq:adjoint-bracket} \nonumber\\
&= \big[\, \mathrm{Ad}_{h}X,\, \mathrm{Ad}_{h}Y \,\big]. 
\end{align}

Next, we would like to compute $\partial_{A A'}\left( \mathrm{Ad}_{h} X \right)$. Now, taking a derivative of $\mathrm{Ad}_{h}X$
\begin{align}
\partial_{A A'} \!\left( h X h^{-1} \right)
&=
(\partial_{A A'} h)\, X\, h^{-1}
+
h\, (\partial_{A A'} X)\, h^{-1}
+
h\, X\, (\partial_{A A'} h^{-1}).
\end{align}
Using the basic identity $\partial_{A A'} h^{-1}
=
- h^{-1} (\partial_{A A'} h)\, h^{-1}$,
this expression can be simplified to
\begin{align}
\partial_{A A'}\!\left( h X h^{-1} \right)
&=
(\partial_{A A'} h)\, X\, h^{-1}
+
h\, (\partial_{A A'} X)\, h^{-1}
-
h\, X\, h^{-1} (\partial_{A A'} h)\, h^{-1}
\nonumber\\
&=
h\!\left(
h^{-1} (\partial_{A A'} h)\, X
+
\partial_{A A'} X
-
X\, h^{-1} (\partial_{A A'} h)
\right) h^{-1}
\nonumber\\
&=
h\!\left(
\partial_{A A'} X
+
\big[ h^{-1} \partial_{A A'} h ,\, X \big]
\right) h^{-1}\nonumber\\&=h(D_{A A'} X)h^{-1}
\end{align}
Here, we defined $D_{A A'} X
=
\partial_{A A'} X
+
\big[ h^{-1} \partial_{A A'} h ,\, X \big]$. Thus, we obtain 
\begin{equation}\label{derivative}
\partial_{A A'} \left( \mathrm{Ad}_{h} X \right)
=
\mathrm{Ad}_{h}\!\left( D_{A A'} X \right),
\qquad
\end{equation}
Applying \eqref{derivative} to $B^{A}$, we obtain
\begin{align}
\partial_{A A'} B^{A}
&=
\partial_{A A'}\!\left(
\mathrm{Ad}_{h}\!\left( \hat{b}\,\mu^{A} - b\,\hat{\mu}^{A} \right)
\right)
\nonumber\\
&=
\mathrm{Ad}_{h}\!\left(
D_{A A'}\!\left( \hat{b}\,\mu^{A} - b\,\hat{\mu}^{A} \right)
\right).
\end{align}
Contracting with $\alpha^{A'}$, we obtain
\begin{equation}
\alpha^{A'} \partial_{A A'} B^{A}
=
\mathrm{Ad}_{h}\!\left(
\alpha^{A'} D_{A A'}\!\left( \hat{b}\,\mu^{A} - b\,\hat{\mu}^{A} \right)
\right).
\end{equation}
Therefore,
\begin{equation}\label{derivative1}
\mathrm{Ad}_{h^{-1}}\!\left(
\alpha^{A'} \partial_{A A'} B^{A}
\right)
=
\alpha^{A'} D_{A A'}\!\left( \hat{b}\,\mu^{A} - b\,\hat{\mu}^{A} \right).
\end{equation}
We now expand the right-hand side using the definition of $D_{A A'}$
\begin{equation}
\alpha^{A'} D_{A A'}\!\left( \hat{b}\,\mu^{A} - b\,\hat{\mu}^{A} \right)
=
\alpha^{A'} \partial_{A A'}\!\left( \hat{b}\,\mu^{A} - b\,\hat{\mu}^{A} \right)
+
\alpha^{A'} \big[ h^{-1} \partial_{A A'} h,\,
\hat{b}\,\mu^{A} - b\,\hat{\mu}^{A} \big].
\end{equation}
Since the spinors $\mu^{A}$ and $\hat{\mu}^{A}$ are constant, the ordinary derivative acts only on $b$ and $\hat{b}$
\begin{equation}
\alpha^{A'} \partial_{A A'}\!\left( \hat{b}\,\mu^{A} - b\,\hat{\mu}^{A} \right)
=
\mu^{A} \alpha^{A'} \partial_{A A'} \hat{b}
-
\hat{\mu}^{A} \alpha^{A'} \partial_{A A'} b.
\end{equation}
For the commutator term, we similarly obtain
\begin{equation}
\alpha^{A'} \big[ h^{-1} \partial_{A A'} h,\,
\hat{b}\,\mu^{A} - b\,\hat{\mu}^{A} \big]
=
\mu^{A} \alpha^{A'} \big[ h^{-1} \partial_{A A'} h,\, \hat{b} \big]
-
\hat{\mu}^{A} \alpha^{A'} \big[ h^{-1} \partial_{A A'} h,\, b \big].
\end{equation}
Altogether, \eqref{derivative1} becomes
\begin{align}
\mathrm{Ad}_{h^{-1}}\!\left(
\alpha^{A'} \partial_{A A'} B^{A}
\right)
&=
\mu^{A} \alpha^{A'} \partial_{A A'} \hat{b}
-
\hat{\mu}^{A} \alpha^{A'} \partial_{A A'} b
\\
&\quad
+
\mu^{A} \alpha^{A'} \big[ h^{-1} \partial_{A A'} h ,\, \hat{b} \big]
-
\hat{\mu}^{A} \alpha^{A'} \big[ h^{-1} \partial_{A A'} h ,\, b \big].
\end{align}
We now rewrite the final two terms using the currents $j$ and $\hat{j}$. By the definition given in \eqref{currents} for $j$ and $\hat{j}$, we seek an explicit decomposition of $\alpha^{A'} h^{-1} \partial_{A A'} h$
in the $\mu$, $\hat{\mu}$ basis. Accordingly, we write
\begin{equation}\label{current decomposition}
\frac{1}{{\langle \alpha\beta\rangle}}\big(\alpha^{A'} h^{-1} \partial_{A A'} h\big)
=
X\,\mu_{A}
+
Y\,\hat{\mu}_{A},
\end{equation}
for some Lie-algebra-valued coefficients $X$ and $Y$ (with no spinor indices). Contracting with $\mu^{A}$ and $\hat{\mu}^{A}$ gives 
\begin{align}\label{current decomposotion 1}
    \frac{1}{{\langle \alpha\beta\rangle}}\big(\alpha^{A'} h^{-1} \partial_{A A'} h\big)
=j \hat{\mu}_A-\hat{j}\mu_A
\end{align}
Then, we find
\begin{align}\label{adjoint derivative}
\mathrm{Ad}_{h^{-1}}\!\left(
\alpha^{A'} \partial_{A A'} B^{A}
\right)
=&
\mu^{A} \alpha^{A'} \partial_{A A'} \hat{b}
-
\hat{\mu}^{A} \alpha^{A'} \partial_{A A'} b+{\langle \alpha\beta\rangle}
\mu^{A}  \big[ j \hat{\mu}_A-\hat{j}\mu_A,\, \hat{b} \big]\\&
-{\langle \alpha\beta\rangle}
\hat{\mu}^{A}  \big[ j \hat{\mu}_A-\hat{j}\mu_A ,\, b \big]\nonumber\\=&\mu^{A} \alpha^{A'} \partial_{A A'} \hat{b}
-
\hat{\mu}^{A} \alpha^{A'} \partial_{A A'} b+{\langle \alpha\beta\rangle}
\big[ j , \hat{b} \big]
-{\langle \alpha\beta\rangle}
 \big[ \hat{j} ,\, b \big].
\end{align}

\section{Proof of Equation \eqref{YB Operator1}} \label{ap C}

We would like to rewrite the operator $\mathcal{K} = (1 - \sigma P)^{-1}(1 - P)$ in terms of $R$. We proceed in two steps: first compute $1 - P$ in terms of $R$, and then compute $1 - \sigma P$ in terms of $R$. Starting from $P = (R - c)(R + c)^{-1},$
we compute
\begin{align}
1 - P
= 1 - \frac{R-c}{R+c}= \frac{2c}{R+c},
\end{align}
\begin{align}
  1 - \sigma P =\frac{\bigl((1 - \sigma)R + (1 + \sigma)c\bigr)}{(R + c)}.
\end{align}
Inverting $1 - \sigma P$, we get 
\begin{align}\label{K OPERATOR}
\mathcal{K}=\frac{2c}{\bigl((1 - \sigma)R + (1 + \sigma)c\bigr)}.
\end{align}
We intend to write
$
\mathcal{N}=(1 - \sigma)R + (1 + \sigma)c
$
as a scalar multiple of $(1 - \eta R)$.  So
$
\mathcal{N}
=
(1 + \sigma)c
\left(
1 + \frac{1 - \sigma}{(1 + \sigma)c} R
\right).
$ We want to identify the expression in the parenthesis with $1 - \eta R$, where $\eta$ is the deformation parameter of the Yang--Baxter sigma model. We therefore define $\eta$ by
\begin{align}
1 - \eta R
&= 1 + \frac{1 - \sigma}{(1 + \sigma)c}\,R
\qquad \Longrightarrow \qquad
\eta = \frac{\sigma - 1}{(\sigma + 1)c}.
\end{align}
Equivalently, we solve for $\sigma$ in terms of $\eta$:
\begin{align}
\eta c
&= \frac{\sigma - 1}{\sigma + 1}
\qquad \Longrightarrow \qquad
\sigma(1 - \eta c) = 1 + \eta c
\qquad \Longrightarrow \qquad
\sigma = \frac{1 + \eta c}{1 - \eta c}.
\end{align}
We find
\begin{align}
\mathcal{N}^{-1}= \frac{1}{(1 + \sigma)c}(1 - \eta R)^{-1}. 
\end{align}
Plugging into \eqref{K OPERATOR}, we obtain
\begin{align}
(1 - \sigma P)^{-1}(1 - P)
&= 2c \cdot \frac{1}{(1 + \sigma)c}(1 - \eta R)^{-1} \nonumber\\
&= \frac{2}{1 + \sigma}(1 - \eta R)^{-1}.
\end{align}
Thus, the identity we require is
\begin{align}
(1 - \sigma P)^{-1}(1 - P)
&= \frac{2}{1 + \sigma}(1 - \eta R)^{-1},
\qquad
\eta = \frac{\sigma - 1}{(\sigma + 1)c}.
\end{align}
Or equivalently (solving for $\sigma$):
\begin{align}
    (1 - \sigma P)^{-1}(1 - P)
=
(1 - \eta c)(1 - \eta R)^{-1},
\qquad
\sigma = \frac{1 + \eta c}{1 - \eta c}.
\end{align}

\section{Gauge transformations for a general operator $\mathcal O$}

\label{subsec:general_O_gauge_boundary}

We first discuss the gauge symmetry of the two-pole boundary condition for a
general linear operator
\begin{equation}
	\mathcal O:\mathfrak g\rightarrow\mathfrak g .
\end{equation}
The Yang--Baxter case is obtained only after the specialization
$\mathcal O=R$.  Keeping $\mathcal O$ general is useful because it makes clear
which algebraic properties are required for the boundary condition to be
preserved by gauge transformations.

The reduced four-dimensional Chern--Simons boundary condition has the form
\begin{equation}
	(\mathcal O-c)A_i\big|_{\zeta=\alpha_+}
	=
	(\mathcal O+c)A_i\big|_{\zeta=\alpha_-},
	\qquad
	i=w,\bar w .
	\label{eq:general_O_BC_CS4}
\end{equation}
Here $\alpha_+$ and $\alpha_-$ denote the two simple poles of the
four-dimensional Chern--Simons one-form.  The finite gauge transformation of
the Chern--Simons connection is
\begin{equation}
	A_i
	\longmapsto
	A_i^u
	=
	u^{-1}\partial_i u
	+
	u^{-1}A_i u .
	\label{eq:finite_gauge_CS4_general_O}
\end{equation}
This is the same convention used later in the parametrisation
\begin{equation}
	A_I
	=
	\hat h^{-1}\mathscr L_I\hat h
	+
	\hat h^{-1}\partial_I\hat h,
	\qquad
	I=w,\bar w .
\end{equation}
Indeed, this parametrisation is obtained by applying a finite gauge
transformation to the meromorphic connection $\mathscr L_I$ with group
element $\hat h^{-1}$.

Let
\begin{equation}
	u_+=u\big|_{\zeta=\alpha_+},
	\qquad
	u_-=u\big|_{\zeta=\alpha_-}.
\end{equation}
After the gauge transformation, the boundary condition becomes
\begin{align}
	&
	(\mathcal O-c)
	\left(
	u_+^{-1}\partial_i u_+
	+
	u_+^{-1}A_i\big|_{\zeta=\alpha_+}u_+
	\right)
	\nonumber\\
	&\hspace{5em}
	=
	(\mathcal O+c)
	\left(
	u_-^{-1}\partial_i u_-
	+
	u_-^{-1}A_i\big|_{\zeta=\alpha_-}u_-
	\right).
	\label{eq:transformed_BC_general_O}
\end{align}
Thus there are two parts to check.  The first is the Maurer--Cartan part
$u_\pm^{-1}\partial_i u_\pm$.  The second is the conjugated gauge-field part
$u_\pm^{-1}A_i u_\pm$.

The original boundary condition \eqref{eq:general_O_BC_CS4} can be solved by
writing the pair of boundary values as
\begin{equation}
	A_i\big|_{\zeta=\alpha_+}
	=
	(\mathcal O+c)x_i,
	\qquad
	A_i\big|_{\zeta=\alpha_-}
	=
	(\mathcal O-c)x_i,
	\label{eq:general_O_allowed_pair_CS4}
\end{equation}
for some $x_i\in\mathfrak g$.  This parametrisation solves the boundary
condition because
\begin{align}
	(\mathcal O-c)A_i\big|_{\zeta=\alpha_+}
	&=
	(\mathcal O-c)(\mathcal O+c)x_i,
	\\
	(\mathcal O+c)A_i\big|_{\zeta=\alpha_-}
	&=
	(\mathcal O+c)(\mathcal O-c)x_i,
\end{align}
and the two operators commute since both are polynomials in $\mathcal O$.
Thus the boundary condition selects the subspace
\begin{equation}
	\mathfrak g_{\mathcal O}
	=
	\left\{
	\big((\mathcal O+c)x,(\mathcal O-c)x\big)
	\ \middle|\ 
	x\in\mathfrak g
	\right\}
	\subset
	\mathfrak g\oplus\mathfrak g .
	\label{eq:gO_subspace}
\end{equation}

For the transformed boundary condition to have the same form, the boundary
values of the gauge transformation must lie in some corresponding group
$G_{\mathcal O}$.  Infinitesimally this means that the Maurer--Cartan pair is
tangent to $\mathfrak g_{\mathcal O}$:
\begin{equation}
	u_+^{-1}\partial_i u_+
	=
	(\mathcal O+c)y_i,
	\qquad
	u_-^{-1}\partial_i u_-
	=
	(\mathcal O-c)y_i,
	\label{eq:MC_pair_general_O}
\end{equation}
for some $y_i\in\mathfrak g$.  Then
\begin{align}
	(\mathcal O-c)u_+^{-1}\partial_i u_+
	&=
	(\mathcal O-c)(\mathcal O+c)y_i,
	\\
	(\mathcal O+c)u_-^{-1}\partial_i u_-
	&=
	(\mathcal O+c)(\mathcal O-c)y_i,
\end{align}
and these two expressions are equal.  Therefore the Maurer--Cartan part of
\eqref{eq:transformed_BC_general_O} preserves the boundary condition.

It remains to check the conjugated gauge-field part.  If the original
boundary pair lies in $\mathfrak g_{\mathcal O}$, then the conjugated pair
will remain in $\mathfrak g_{\mathcal O}$ provided
$\mathfrak g_{\mathcal O}$ is stable under the adjoint action of
$G_{\mathcal O}$.  This is guaranteed if $\mathfrak g_{\mathcal O}$ is a Lie
subalgebra of $\mathfrak g\oplus\mathfrak g$.  We therefore compute the
closure condition.

Take two elements of $\mathfrak g_{\mathcal O}$,
\begin{equation}
	X_{\mathcal O}
	=
	\big((\mathcal O+c)X,(\mathcal O-c)X\big),
	\qquad
	Y_{\mathcal O}
	=
	\big((\mathcal O+c)Y,(\mathcal O-c)Y\big).
\end{equation}
Their commutator in $\mathfrak g\oplus\mathfrak g$ is
\begin{align}
	[X_{\mathcal O},Y_{\mathcal O}]
	=
	\Big(
	[(\mathcal O+c)X,(\mathcal O+c)Y],
	[(\mathcal O-c)X,(\mathcal O-c)Y]
	\Big).
	\label{eq:gO_commutator_general_O}
\end{align}
For closure, this must again be of the form
\begin{equation}
	\big((\mathcal O+c)Z,(\mathcal O-c)Z\big)
\end{equation}
for some $Z\in\mathfrak g$.  Hence we require
\begin{align}
	[(\mathcal O+c)X,(\mathcal O+c)Y]
	&=
	(\mathcal O+c)Z,
	\label{eq:closure_plus_general_O}
	\\
	[(\mathcal O-c)X,(\mathcal O-c)Y]
	&=
	(\mathcal O-c)Z.
	\label{eq:closure_minus_general_O}
\end{align}

We now expand both components explicitly.  The first component gives
\begin{align}
	[(\mathcal O+c)X,(\mathcal O+c)Y]
	&=
	[\mathcal OX+cX,\mathcal OY+cY]
	\nonumber\\
	&=
	[\mathcal OX,\mathcal OY]
	+
	c[\mathcal OX,Y]
	+
	c[X,\mathcal OY]
	+
	c^2[X,Y].
	\label{eq:plus_expand_general_O}
\end{align}
The second component gives
\begin{align}
	[(\mathcal O-c)X,(\mathcal O-c)Y]
	&=
	[\mathcal OX-cX,\mathcal OY-cY]
	\nonumber\\
	&=
	[\mathcal OX,\mathcal OY]
	-
	c[\mathcal OX,Y]
	-
	c[X,\mathcal OY]
	+
	c^2[X,Y].
	\label{eq:minus_expand_general_O}
\end{align}
The terms proportional to $c$ fix $Z$ uniquely as
\begin{equation}
	Z
	=
	[\mathcal OX,Y]+[X,\mathcal OY].
	\label{eq:Z_general_O}
\end{equation}
Substituting this into $(\mathcal O\pm c)Z$, we get
\begin{align}
	(\mathcal O+c)Z
	&=
	\mathcal O
	\left(
	[\mathcal OX,Y]+[X,\mathcal OY]
	\right)
	\nonumber\\
	&\quad
	+
	c
	\left(
	[\mathcal OX,Y]+[X,\mathcal OY]
	\right),
	\\
	(\mathcal O-c)Z
	&=
	\mathcal O
	\left(
	[\mathcal OX,Y]+[X,\mathcal OY]
	\right)
	\nonumber\\
	&\quad
	-
	c
	\left(
	[\mathcal OX,Y]+[X,\mathcal OY]
	\right).
\end{align}
Comparing with \eqref{eq:plus_expand_general_O} and
\eqref{eq:minus_expand_general_O}, the closure condition is
\begin{equation}
	[\mathcal OX,\mathcal OY]
	+
	c^2[X,Y]
	=
	\mathcal O
	\left(
	[\mathcal OX,Y]+[X,\mathcal OY]
	\right).
\end{equation}
Equivalently,
\begin{equation}
	[\mathcal OX,\mathcal OY]
	-
	\mathcal O
	\left(
	[\mathcal OX,Y]+[X,\mathcal OY]
	\right)
	=
	-c^2[X,Y].
	\label{eq:general_O_mCYBE}
\end{equation}
Thus $\mathfrak g_{\mathcal O}$ is a Lie subalgebra precisely when
$\mathcal O$ satisfies the modified classical Yang--Baxter equation.

There is also an isotropy requirement from the boundary term in the
Chern--Simons variation.  The relevant pairing on
$\mathfrak g\oplus\mathfrak g$ is the difference of the two trace pairings.
For two elements of $\mathfrak g_{\mathcal O}$, this gives
\begin{align}
	&
	\Tr\left[
	(\mathcal O+c)X\,(\mathcal O+c)Y
	\right]
	-
	\Tr\left[
	(\mathcal O-c)X\,(\mathcal O-c)Y
	\right]
	\nonumber\\
	&\hspace{3em}
	=
	2c\,
	\Tr\left[
	\mathcal OX\,Y
	+
	X\,\mathcal OY
	\right].
\end{align}
This vanishes if
\begin{equation}
	\mathcal O^t=-\mathcal O,
	\label{eq:general_O_skew}
\end{equation}
where the transpose is defined with respect to the invariant trace pairing.
Thus the boundary condition is compatible with the Chern--Simons variational
principle and finite gauge transformations when
\begin{equation}
	\mathcal O^t=-\mathcal O,
	\qquad
	[\mathcal OX,\mathcal OY]
	-
	\mathcal O
	\left(
	[\mathcal OX,Y]+[X,\mathcal OY]
	\right)
	=
	-c^2[X,Y].
	\label{eq:general_O_conditions}
\end{equation}
If a general operator $\mathcal O$ does not satisfy these conditions, the
boundary values do not define a stable boundary algebra.  In that case, the boundary
condition is not gauge invariant.

The Yang--Baxter case used in the main construction is obtained by setting
\begin{equation}
	\mathcal O=R.
\end{equation}
Then \eqref{eq:general_O_mCYBE} becomes the modified classical Yang--Baxter
equation for $R$, and the closure identity may be written as
\begin{equation}
	[(R\pm c)X,(R\pm c)Y]
	=
	(R\pm c)
	\left(
	[RX,Y]+[X,RY]
	\right).
	\label{eq:Rpm_identity}
\end{equation}
This is the form used in the finite gauge-transformation argument.  It shows
that the images of $R+c$ and $R-c$ are Lie subalgebras, and therefore the
finite boundary values of the allowed gauge transformations form the
corresponding Yang--Baxter subgroup.

\subsection{The same general operator in six dimensions}
\label{subsec:general_O_six_dimensional}

We now explain how the same algebraic condition appears before symmetry
reduction, directly in the six-dimensional twistor theory.  The
six-dimensional boundary conditions for a general operator $\mathcal O$ are
\begin{align}
	(\mathcal O-c)[\mathcal A\mu]\big|_{\pi=\alpha}
	&=
	\sigma
	\frac{\langle\alpha\beta\rangle}
	     {\langle\tilde\alpha\beta\rangle}
	(\mathcal O+c)[\mathcal A\mu]\big|_{\pi=\tilde\alpha},
	\label{eq:6d_mu_general_O}
	\\
	(\mathcal O-c)[\mathcal A\hat\mu]\big|_{\pi=\alpha}
	&=
	\sigma^{-1}
	\frac{\langle\alpha\beta\rangle}
	     {\langle\tilde\alpha\beta\rangle}
	(\mathcal O+c)[\mathcal A\hat\mu]\big|_{\pi=\tilde\alpha}.
	\label{eq:6d_hatmu_general_O}
\end{align}
We take the same choice of $\sigma$ as in the Yang--Baxter construction, so
that after symmetry reduction the spinor factors convert
\eqref{eq:6d_mu_general_O} and \eqref{eq:6d_hatmu_general_O} into the
four-dimensional boundary condition discussed above.

The algebraic part of the six-dimensional condition has the same structure as
in CS$_4$.  In the $\mu$ component, the allowed boundary values may be written
as
\begin{equation}
	[\mathcal A\mu]\big|_{\pi=\alpha}
	=
	\langle\alpha\gamma\rangle(\mathcal O+c)x,
	\qquad
	[\mathcal A\mu]\big|_{\pi=\tilde\alpha}
	=
	\langle\tilde\alpha\gamma\rangle(\mathcal O-c)x.
	\label{eq:6d_mu_allowed_general_O}
\end{equation}
Similarly, the hatted component may be written as
\begin{equation}
	[\mathcal A\hat\mu]\big|_{\pi=\alpha}
	=
	\langle\alpha\hat\gamma\rangle(\mathcal O+c)y,
	\qquad
	[\mathcal A\hat\mu]\big|_{\pi=\tilde\alpha}
	=
	\langle\tilde\alpha\hat\gamma\rangle(\mathcal O-c)y.
	\label{eq:6d_hatmu_allowed_general_O}
\end{equation}
The spinor factors only keep track of which component is being reduced to
$A_w$ or $A_{\bar w}$.  The Lie-algebraic content is still the pair
\begin{equation}
	\big((\mathcal O+c)x,(\mathcal O-c)x\big).
\end{equation}
Therefore the conjugation part of a six-dimensional gauge transformation
preserves the algebraic form of the boundary values if and only if
$\mathfrak g_{\mathcal O}$ is a Lie subalgebra.  As shown above, this is
equivalent to the modified classical Yang--Baxter equation
\eqref{eq:general_O_mCYBE}.

The new feature in six dimensions is the derivative part of the gauge
transformation.  The six-dimensional connection transforms as
\begin{equation}
	\mathcal A
	\longmapsto
	\mathcal A^u
	=
	u^{-1}\mathcal A u
	+
	u^{-1}\bar\partial u.
	\label{eq:6d_gauge_general_O}
\end{equation}
For the components along $\mathbb E^4$, this gives
\begin{equation}
	\mathcal A_A
	\longmapsto
	\mathcal A_A^u
	=
	u^{-1}\mathcal A_Au
	+
	u^{-1}\pi^{A'}\partial_{AA'}u.
	\label{eq:6d_component_gauge_general_O}
\end{equation}
Therefore
\begin{align}
	[\mathcal A^u\mu]\big|_{\pi=\alpha}
	=
	&
	u_\alpha^{-1}
	[\mathcal A\mu]\big|_{\pi=\alpha}
	u_\alpha
	\nonumber\\
	&
	+
	\mu^A\alpha^{A'}
	u_\alpha^{-1}\partial_{AA'}u_\alpha,
	\label{eq:6d_Amu_transform_alpha_general_O}
\end{align}
and similarly at $\pi=\tilde\alpha$.  The first term is the conjugation part,
controlled by the boundary algebra above.  The second term contains
derivatives along the four spacetime directions before symmetry reduction.

To see the effect of the derivative term, decompose a primed spinor in the
$\gamma,\hat\gamma$ basis:
\begin{equation}
	X^{A'}
	=
	\langle X\hat\gamma\rangle\gamma^{A'}
	-
	\langle X\gamma\rangle\hat\gamma^{A'}.
	\label{eq:spinor_decomposition_general_O}
\end{equation}
For the $\mu$ component, this gives
\begin{align}
	\mu^A\alpha^{A'}
	u_\alpha^{-1}\partial_{AA'}u_\alpha
	=
	&
	\langle\alpha\hat\gamma\rangle\,
	\mu^A\gamma^{A'}
	u_\alpha^{-1}\partial_{AA'}u_\alpha
	\nonumber\\
	&
	-
	\langle\alpha\gamma\rangle\,
	\mu^A\hat\gamma^{A'}
	u_\alpha^{-1}\partial_{AA'}u_\alpha .
	\label{eq:mu_derivative_decomp_general_O}
\end{align}
The term with $\hat\gamma^{A'}$ is the piece that becomes the $w$ derivative
after symmetry reduction.  The term with $\gamma^{A'}$ points along the
direction removed by the reduction.  In reduction coordinates,
\begin{equation}
	\mu^A\gamma^{A'}\partial_{AA'}=\partial_{\bar z},
	\qquad
	\mu^A\hat\gamma^{A'}\partial_{AA'}=-\partial_w,
\end{equation}
up to the chosen orientation convention for $w$.  The $\partial_w$ part is
the derivative term seen by the reduced CS$_4$ connection.  The
$\partial_{\bar z}$ part has no counterpart in the CS$_4$ boundary condition.
Thus preservation of the six-dimensional $\mu$ boundary condition requires
\begin{equation}
	\mu^A\gamma^{A'}\partial_{AA'}u\big|_{\pi=\alpha,\tilde\alpha}=0.
	\label{eq:semilocal_mu_general_O}
\end{equation}

Repeating the same analysis for the hatted component gives the complementary
condition
\begin{equation}
	\hat\mu^A\hat\gamma^{A'}\partial_{AA'}u\big|_{\pi=\alpha,\tilde\alpha}=0.
	\label{eq:semilocal_hatmu_general_O}
\end{equation}
In the same reduction coordinates,
\begin{equation}
	\hat\mu^A\hat\gamma^{A'}\partial_{AA'}=\partial_z,
	\qquad
	\hat\mu^A\gamma^{A'}\partial_{AA'}=\partial_{\bar w}.
\end{equation}
Hence the allowed six-dimensional gauge transformations must obey
\begin{equation}
	\partial_{\bar z}u\big|_{\pi=\alpha,\tilde\alpha}=0,
	\qquad
	\partial_z u\big|_{\pi=\alpha,\tilde\alpha}=0.
	\label{eq:semilocal_conditions_general_O}
\end{equation}

This is the origin of semi-locality.  The algebraic part of the symmetry is
the same as in CS$_4$: it is governed by the subgroup generated by
$\mathfrak g_{\mathcal O}$.  However, before symmetry reduction, the gauge
parameter also appears through derivatives along $z$ and $\bar z$.  These
directions are removed in the reduction, and their derivative contributions
must vanish for the six-dimensional boundary condition to be preserved.
Thus the symmetry is not a fully local gauge symmetry of the four-dimensional
IFT.  It is local only along the directions which survive the reduction, while
being constrained along the removed directions.

For a general operator $\mathcal O$, the six-dimensional boundary condition
therefore has a semi-local symmetry only if the following conditions hold:
\begin{equation}
	\mathcal O^t=-\mathcal O,
\end{equation}
\begin{equation}
	[\mathcal OX,\mathcal OY]
	-
	\mathcal O
	\left(
	[\mathcal OX,Y]+[X,\mathcal OY]
	\right)
	=
	-c^2[X,Y],
\end{equation}
and
\begin{equation}
	\mu^A\gamma^{A'}\partial_{AA'}u=0,
	\qquad
	\hat\mu^A\hat\gamma^{A'}\partial_{AA'}u=0
\end{equation}
at the relevant pole values.  The first two conditions are algebraic
conditions on $\mathcal O$.  The last two conditions are differential
conditions on the six-dimensional gauge parameter.  If $\mathcal O$ is not of
Yang--Baxter type, the algebraic boundary values do not close; if the
differential conditions are not imposed, the six-dimensional derivative terms
spoil the boundary condition.

\bibliographystyle{ytphys}  % or alpha, unsrt, etc.
\bibliography{twistor}

\end{document}